\documentstyle[aps,pra,epsfig,twocolumn]{revtex}

\def\be{\begin{equation}}
\def\ee{\end{equation}}
\def\bea{\begin{eqnarray}}
\def\eea{\end{eqnarray}}

\begin{document}
\draft

\title{Quantum computing with neutral atoms}

\author{
H.-J.~Briegel,$^{1}$  T.~Calarco,$^{2,3}$ D.~Jaksch,$^{2}$ J.I.~Cirac,$^2$ and 
P.~Zoller$^2$
}

\address{$^{1}$Sektion Physik, Ludwig-Maximilians-Universit\"at M\"unchen,
D-80333 M\"unchen, Germany \\
$^{2}$Institut f\"ur Theoretische Physik, Universit\"at Innsbruck,
Technikerstrasse 25, A-6020 Innsbruck, Austria\\
$^3$ECT*, European Centre for Theoretical Studies in Nuclear Physics
and Related Areas \\
Villa Tambosi, Strada delle Tabarelle 286, Villazzano (Trento), Italy 38050}

\maketitle

\begin{abstract}
We develop a method to entangle neutral atoms using cold controlled 
collisions. We analyze this method in two particular set-ups: optical lattices and 
magnetic micro-traps. Both offer the possibility of performing
certain multi-particle operations in parallel. Using this fact, we show how
to implement efficient quantum error correction and schemes for fault-tolerant computing. 
\end{abstract}

\section{Introduction}

Entanglement is one of the most intriguing features of Quantum
Mechanics. However, there are very few physical systems in which
entanglement can be systematically studied in a controlled way. Those
systems include ion-traps \cite{cirac95,winelandgroup1,winelandgroup2,winelandgroup3,winelandgroup4,steane1,steane2,lanlions}, 
cavity QED
\cite{turchette95,maitre97,hagley97,Pellizzari96,quantumnetworks1,quantumnetworks1b,quantumnetworks2,quantumnetworks3},
photons \cite{Aspect1,Aspect2,Aspect3,Cabrillo,photons1,photons2,photons3,photons4,photons5}, 
and molecules in the context of NMR \cite
{cory97,gershenfeld97,othernmrgroups1,othernmrgroups2} (see
\cite{BraunsteinNMR} however). Very recently, we have identified a new
way of entangling particles by using {\it cold controlled collisions}
with which one could study experimentally basic issues of quantum
information theory \cite{jaksch98}. Given the impressive experimental
advances made so far in the fields of neutral atom trapping and cooling
\cite{cooling-trapping,NOBEL1,NOBEL2,NOBEL3}, and in the studies of Bose Einstein
condensation (BEC) of ultracold gases
\cite{BEC951,BEC952,BEC953,hall981,hall983,stamper-kurn98}, that
proposal opens a new perspective to several experimental groups who so
far have concentrated their efforts in other fields of Atomic Physics. 

In the present paper, we build upon the work in \cite{jaksch98} and
explore the idea of using atomic controlled cold collisions for
entangling neutral atoms in optical lattices (see also \cite{brennen})
and in arrays of magnetic micro-traps. We show how to perform two-qubit
gate operations with those systems obtaining very high fidelities. We 
propose a variety of experiments to entangle particles using state-of-the-art 
technology. We also concentrate on the unique
possibilities that these set-ups offer to perform multi-particle
entanglement operations in parallel
\cite{parallelqc1,parallelqc2,parallelqc3,parallelqc4}. 
Using such parallelism, we show how to implement
efficient error correction \cite{shor-code,QECC1,QECC2,laflamme96,QECC3,QECC4,QECC5,QECC-review}
and fault-tolerant quantum computation schemes 
\cite{fault-tolerant1,fault-tolerant2,knill96,othersthanknill1,othersthanknill2,othersthanknill3,gottesman98,fault-tolerant-review}. 

The paper is organized as follows. In Sec.~II we discuss the use of
ultracold collisions as a mechanism for entangling neutral atoms. Such
collisions can be brought about by either moving the potentials in
certain spatial directions or by modifying the shape of the trapping
potentials. In Sec.~III we describe two systems in which such operations
can be implemented. These are optical lattices \cite
{brennen,opticallattices2,opticallattices3,opticallattices4,opticallattices5} and magnetic microtraps
\cite{magneticmicrotraps1,magneticmicrotraps2,magneticmicrotraps3,magneticmicrotraps4,magneticmicrotraps5} 
both of which have been studied
experimentally in detail in the past. In Sec. IV we describe a class of
multi-particle entanglement operations that can be realized in these
systems (we concentrate here on optical lattices). The usefulness of
such operations for quantum computing depends on certain conditions that
need to be satisfied in an experiment. Among these conditions, the
filling problem, i.e. how to fill the potentials with regular patterns
of atoms, is most outstanding. We discuss these matters and show that
even under present-day experimental conditions, very interesting
entanglement studies could be performed. Section V summarizes the main
results and discusses their relevance for future research.

\section{Entanglement of atoms via cold controlled collisions}
\label{EntviaCCC}

In this Section, we consider two bosonic neutral atoms with two internal states trapped by 
conservative potentials and cooled to the motional ground states. Initially these two 
particles are sufficiently far apart so that they do not interact with each other. 
We then assume the shape of the potentials to be varied in a way that depends on
the internal state of the atoms so that the two particles come close to each 
other if they are in certain internal states. As we will show, this
can be done e.g.~by moving the center position of the trapping potentials 
state selectively, or by switching off a potential barrier between the two atoms
for one of the two internal states. In both cases the particles will interact 
via $s$-wave scattering with each other in a coherent way when they are close 
to each other. After the interaction has taken place the particles are restored 
to their initial position. In this way one can implement 
conditional dynamics and realize a fundamental two-qubit gate.

Note that we are dealing with bosons.
Therefore, we have to use symmetrized wave functions for describing the two particles.
It will turn out that if the center positions of the trapping potentials are moved
state selectively, particles in the same internal state will always be so far apart that their 
wave functions never overlap. Thus, we will not care about the symmetrization in this case. 
On the other hand, if the potential barrier is switched off for one internal state, particles in the same 
internal state will come close to each other and symmetrizing the wave function is essential.

\subsection{Hamiltonian}

Here we deal with the interaction Hamiltonian of 
two neutral atoms $1$ and $2$ with internal states $|a\rangle_{1,2}$ and
$|b\rangle_{1,2}$ trapped by conservative potentials $V^{\beta_i}({\bf x}_i,t)$
whose functional dependence on the coordinate ${\bf x}_i$, with $i=1,2$ the particle
index, depends on the 
internal state of the particle $\beta_{1,2}=a,b$. Initially, the two particles are
in the ground state of the trapping potentials and the centers of the two
potential wells are sufficiently far apart so that the particles do not
interact. Then the form of the potential wells is changed such that there
is some overlap of the wave functions of the two atoms, and 
the particles will interact with each other. This interaction between the 
atoms in two given internal states $\beta_1$ and $\beta_2$ can be described by a 
contact potential 
\be
u^{\beta_1\beta_2}({\bf x}_1-{\bf x}_2)=\frac{4 \pi a_s^{\beta_1\beta_2} \hbar^2}{m} 
\delta^3 ({\bf x}_1-{\bf x}_2),
\ee
where $a_s^{\beta_1\beta_2}$ is the $s$-wave scattering length for the corresponding
internal states describing elastic collisions and $m$ is the mass of the particles. This 
zero energy $s$-wave scattering approximation will be valid as long as we assume that 
$v_{\rm osc}$, the rms velocity of the atoms in the vibrational ground state, 
approximately given by $v_{\rm osc} \approx a_0\omega$, is sufficiently small \cite{note}. 
Here $a_0$ is the size of the ground state of the trap potential, and $\omega$ is the first excitation 
frequency. Thus we can describe the evolution of the system by the Hamiltonian
\be
\label{HamilTot}
H=\sum_{\beta_1,\beta_2} H^{\beta_1\beta_2} \otimes 
|\beta_1\rangle_1\langle \beta_1| \otimes |\beta_2\rangle_2\langle \beta_2|,
\ee
where
\be
\label{Hamil}
H^{\beta_1\beta_2}= \sum_{i=1,2} \left[ \frac{({\bf p}_i)^2}{2m} + 
V^{\beta_i}\left({\bf x}_i,t\right) \right] + u^{\beta_1 \beta_2}({\bf x}_1-{\bf x}_2).
\ee
Here ${\bf p}_i$ is the momentum operator.

\subsubsection{Interaction in perturbation theory}

We want to treat the interaction term in the Hamiltonian Eq.~(\ref{Hamil})
perturbatively. For particles in two different internal states
$\beta_1 \neq \beta_2$ we find the correction to the energy due to the
interaction as
\be
\label{deltaE}
\Delta E^{\beta_1\beta_2}(t)= \frac{4\pi a^{ab}_s\hbar^2}{m}\int d{\bf x}
\prod_{i}
\left| \psi^{\beta_i}_i \left({\bf x},t\right)\right|^2,
\ee
where $\psi^{\beta_i}_i \left({\bf x},t\right)$ is the normalized one-particle 
wave function of 
particle $i$ in internal state $\beta_i$ in the time dependent potential 
$V^{\beta_i}\left({\bf x},t\right)$.
If the particles are in the same internal state
$\beta_1=\beta_2=\beta$, we have to
account for the Bose statistics i.e.~use the properly normalized symmetrized two-particle
wave function for calculating the energy shift. We therefore find
\be
\label{deltaEbos}
\Delta E^{\beta\beta}(t)= \frac{8\pi a^{\beta\beta}_s\hbar^2}{m (1+|\alpha|^2)}
\int dx 
\prod_{i}\left|\psi^{\beta}_i \left({\bf x},t\right)\right|^2,
\ee
where 
\be
\alpha=\int dx \left(\psi^{\beta}_1 \left({\bf x},t\right)\right)^*
\psi^{\beta}_2 \left({\bf x},t\right).
\ee
For general $\beta_1$, $\beta_2$ we find the phase accumulated due to the interaction 
in the time interval $[-\tau,\tau]$ by
\be
\label{phicol}
\phi^{\beta_1\beta_2} = \frac {1}{\hbar} \int_{-\tau}^\tau \!dt \, \Delta
E^{\beta_1\beta_2}(t).
\ee

\subsection{Moving potentials}
\label{Movpot}

One way of controlling the interaction between the particles is to move the
center position of the potentials $V^{\beta_i} \left({\bf x}_i,t\right)=
V\left({\bf x}_i-\bar {\bf x}_i^{\beta_i}(t)\right)$ towards each other
in a state-dependent way while leaving the shape of the potential unchanged.
By moving the potential we get two kinds of phase shifts. A kinetic phase
which is a single-particle phase due to the kinetic energy of the particles
and an interaction phase due to coherent interactions between two atoms.
First we will define these two phases for general trapping potentials and afterwards
specialize them to moving harmonic potentials. Finally, we will show how conditional
dynamics can be realized.

\subsubsection{Kinetic phase}
\label{KinPhas}

First we want to consider a single atom in internal state
$|\beta\rangle$ trapped in the instantaneous ground state $\psi_0$ of a
moving potential well $V({\bf x}-\bar {\bf x}^\beta(t))$.
The center position of the potential is moved along a trajectory $\bar
{\bf x}^\beta(t)$.
Ideally, we want the atom to remain in the ground state of its trapping 
potential and to preserve its internal state during the motion. This corresponds to the 
transformation from $t=-\tau$ to $t=\tau$
\be
\label{transf1}
\psi_0[{\bf x} -\bar {\bf x}^\beta(-\tau)] \rightarrow
e^{-i\phi^\beta} \psi_0[{\bf x}-\bar {\bf x}^\beta(\tau)],
\ee
where the atom remains in the ground state of the trapping 
potential and preserves its internal state. Transformation
(\ref{transf1}) can be realized in the {\em adiabatic limit} \cite{GalPas},
where we move the potentials so that the atoms remain in the instantaneous
motional ground state. Adiabaticity requires $|\dot{\bar {\bf x}}^\beta (t)| \ll
v_{\rm osc}$ for all times $t$. The phase $\phi^\beta$ can be easily
calculated in the limit $|\ddot{\bar {\bf x}}^\beta(t)| \ll v_{\rm osc}/\tau$.
We find the {\em kinetic phase}
\be
\phi^{\beta} = \frac{m}{2 \hbar} \int_{-\tau}^\tau \!dt \, \left(\dot{\bar {\bf x}}^\beta(t)\right)^2.
\ee
 
\subsubsection{Interaction phase\label{sect:phinterlattice}}

Let us now consider two particles $i=1,2$ in different internal states 
$|\beta_i\rangle_i$ trapped in the ground states of two moving potentials. 
Initially, at time $t=-\tau$, 
these wells are centered at positions $\bar {\bf x}_i$, 
sufficiently far apart (distance $d=\bar {\bf x}_1 -\bar {\bf x}_2$)
so that the particles do not interact. The positions of the potentials are moved 
along trajectories $\bar {\bf x}^{\beta_i}_i(t)$ so that the wave packets of the 
atoms overlap for certain time, until finally they are restored to the 
initial position at the time $t=\tau$.
We assume that: (i) $|\dot{\bar {\bf x}}^{\beta_i}_i(t)| \ll v_{\rm osc}$
(adiabatic condition) so that the particles remain in the ground states
of the moving trapping potentials; (ii) The interaction can be treated 
perturbatively, where $|\Delta E^{\beta_1\beta_2}(t)|\ll \hbar \omega$ so 
that no sloshing motion is excited.
In that case, we realize the transformation
\begin{eqnarray}
\label{transf}
&&\psi_0({\bf x}_1-\bar {\bf x}_1) \psi_0({\bf x}_2-\bar 
{\bf x}_2) 
\rightarrow \nonumber \\
&&\qquad e^{-i\phi} \psi_0({\bf x}_1-\bar {\bf x}_1) \psi_0({\bf x}_2-\bar 
{\bf x}_2),
\end{eqnarray}
where
$\phi=\phi^{\beta_1} + \phi^{\beta_2} + \phi^{\beta_1\beta_2}$ with the 
{\em collisional phase} $\phi^{\beta_1\beta_2}$ defined in 
Eq.~(\ref{phicol}).

\subsubsection{Moving harmonic potentials}

Here we specialize to harmonic trapping potentials.
The wave function $\psi^{\beta_i}_i({\bf x},t)$ of a particle in a moving harmonic potential 
can be found analytically. In the Appendix {\ref{MovHarm}} we show that when we start
to move the harmonic potential at time $-\tau$ with the particle in its motional ground
state and stop to move the potential at time $\tau$, the condition for the particle to end up 
in the motional ground state at $\tau$ is given by
\be
\label{adia}
\left| \int_{-\tau}^\tau \dot{\bar {\bf x}}^{\beta_i}_i(t') e^{i\omega t'} dt' \right| \ll a_0.
\ee
This condition is weaker than the condition $|\dot{\bar {\bf x}}^{\beta_i}_i(t)| \ll v_{\rm osc}$
for adiabaticity, and means that the particle need not be in
the instantaneous ground state of the moving potential at all times, but only at the
final time. 
The kinetic phases can be found exactly (cf. Eq.~(\ref{phiex})). 
If $|\Delta E^{\beta_1\beta_2}(t)|\ll \hbar \omega$ is satisfied, the interaction phase can be 
found by Eq.~(\ref{phicol}) since the  $\psi^{\beta_i}_i \left({\bf x},t\right)$  are  known. 
It is also possible to generalize these results to the case in which the trap 
frequency changes with time \cite{Shlyap}.

\subsubsection{Implementation of conditional dynamics}

\begin{figure}[tbp]
\epsfig{file=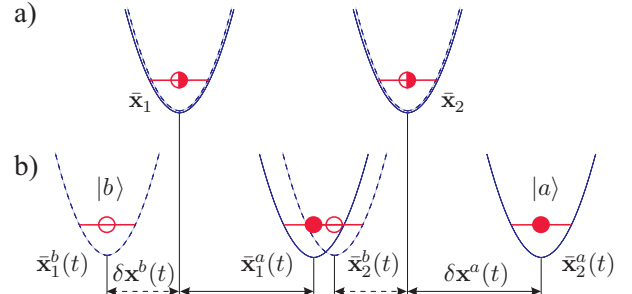,width=8cm}
\caption{Configurations at times $\pm \tau$ (a) and at $t$ (b). The 
solid (dashed) curves show the potentials for particles in the internal state 
$|a\rangle$ ($|b\rangle$). Center positions $\bar
{\bf x}^{\beta_i}_i(t)$ and displacements $\delta {\bf x}^{\beta_i}(t)$ as defined in the
text. \label{ConfiFig}}
\end{figure}

Let us now assume that we can design the potentials such that atoms in the internal 
state $|\beta_i\rangle_i$ experience a potential $V^{\beta_i}({\bf x}_i,t)=V({\bf x}_i-\bar
{\bf x}^{\beta_i}_i(t))$ which is
initially ($t=-\tau$) centered at position $\bar {\bf x}_i$. We assume that we can
move the centers of the potentials as follows: $\bar
{\bf x}^{\beta_i}_i(t)=\bar {\bf x}_i+ \delta {\bf x}^{\beta_i}(t)$. As shown in 
Fig.\ \ref{ConfiFig} the trajectories 
$\delta {\bf x}^{\beta_i}(t)$ are chosen in such a way that 
$\delta {\bf x}^{\beta_i}(-\tau)=\delta {\bf x}^{\beta_i}(\tau)=0$ 
and the first atom collides with the second one only if they are in states $|a\rangle$
and $|b\rangle$, respectively ($|\bar {\bf x}_1^b(t)-\bar {\bf x}_2^a(t)|\gg a_0$ $\forall t$).
This choice is motivated by the physical implementation considered in 
Sec.~\ref{PhysImpLatt}. The fact that $\bar {\bf x}_i$ does not 
depend on the internal atomic state and the shape of the two potentials
is the same at times $\pm \tau$ allows one to easily change the
internal state at times $t=\pm \tau$ by applying laser pulses. If the conditions
stated above are fulfilled, depending on the initial internal atomic states we 
have:
\bea
\label{gate}
|a\rangle_1 |a\rangle_2 &\rightarrow& 
  e^{-i2\phi^a} |a\rangle_1 |a\rangle_2,\nonumber\\
|a\rangle_1 |b\rangle_2 &\rightarrow& 
  e^{-i(\phi^a+\phi^b+\phi^{ab})} |a\rangle_1 |b\rangle_2,\nonumber\\
|b\rangle_1 |a\rangle_2 &\rightarrow& 
  e^{-i(\phi^a+\phi^b)} |b\rangle_1 |a\rangle_2,\nonumber\\
|b\rangle_1 |b\rangle_2 &\rightarrow& 
  e^{-i2\phi^b} |b\rangle_1 |b\rangle_2,
\eea
where the motional states remain unchanged. 
The kinetic phases $\phi^\beta$ and the collisional phase
$\phi^{ab}$ can be calculated as stated above. 
We emphasize that the $\phi^\beta$ are (trivial)
one-particle phases that, as long as they are known, can always be 
incorporated in the definition of the states $|a\rangle$ and $|b\rangle$.
This realizes a fundamental two-qubit quantum gate for certain values
of $\phi^{ab}$, e.g.~$\phi^{ab}=\pi$. 

\subsection{\label{section:switching}Switching potentials}

The interaction between the particles can be controlled also in another way, 
for example by changing with time the shape of the potentials depending on the 
particles' internal states. Different regimes for the
time-dependence of the potential are possible. 
The two limits of extremely slow (adiabatic)
or extremely fast (sudden) potential changes are both interesting and lead to
peculiar schemes. Here we will analyze the latter case. 
We consider two atoms initially trapped in two displaced wells. At a certain
time the barrier between the wells is suddenly removed in a selective way
for atoms in state
$|b\rangle$, whereas it remains unchanged for atoms in state $|a\rangle$. The
atoms are allowed to oscillate for some time, and then the barrier is raised
again suddenly such as to trap them back at the original positions. During the
process they will acquire both a kinematic phase due to the oscillations within
their respective wells, and an interaction phase due to the collision.
We will calculate such quantities and look for the optimal switching time
required in order to maximize the fidelity for a quantum gate relying on this
scheme, which we will estimate quantitatively
for the relevant physical example in Sec. \ref{Twogatemagn}.

\subsubsection{Kinematic phase}

Let us first consider the time-independent problem of
an atom subject to a three-dimensional potential whose functional 
form along $x$ depends on the internal atomic state $\beta=a,b$:
\be
\label{Switch1part}
V^\beta({\bf x})=v^\beta(x)+v_\perp(y)+v_\perp(z).
\ee
Here the $v$'s are single-well trapping potentials, and $v^a$, $v^b$ 
are centered around $x=x_0$, $x=0$, respectively. 
We assume that the atom is initially prepared in
the motional state $|\Psi_+\rangle$, where 
\be
\label{Psi+}
\langle{\bf x}|\Psi_+\rangle\equiv\Psi_+({\bf x})=
\psi_+(x)\psi_\perp(y)\psi_\perp(z)
\ee
and $\psi_+$, $\psi_\perp$ are the ground-state wave functions of $v^a$,
$v_\perp$  with eigenvalues $E^a$, $E_\perp$ respectively. 
Thus $\Psi_+({\bf x})$ is peaked around the position 
${\bf x}_0\equiv (x_0,0,0)$, coinciding with the center of $V^a({\bf x})$ but 
displaced from the one of $V^b({\bf x})$. Therefore, if the atom is in internal
state $|a\rangle$, its motional state after
a time $t$ will be unchanged up to a phase 
$\phi^a=(E^a+2E_\perp)t/\hbar$. 
If it is instead in state $|b\rangle$, it
will start oscillating within the well, thus picking up a different phase $\phi^b$ 
due to the kinematical evolution, and possibly coming back at the initial
position after some time.

\subsubsection{Interaction phase\label{section:phinter}}

We now consider two atoms 1 and 2 initially (at $t=0$)
prepared in the motional states $|\Psi_+\rangle$ 
and $|\Psi_-\rangle$, the latter being defined as in Eq.~(\ref{Psi+}) but with
$\psi_-(x)\equiv\psi_+(-x)$ replacing $\psi_+(x)$.
We assume that the particles are subject to the potentials 
$\sum_{\varsigma=+,-}\Theta(\varsigma x)V^{\beta_i}(\varsigma {\bf x}_i)$,
where $\Theta$ denotes the step function. 
If any one of them is 
in state $|b\rangle$, for $t>0$ it will start oscillating within the well, 
eventually interacting with the other one.
If $v_\perp$ is much steeper than $v^\beta$, then the probability of transversal
excitations can be neglected, {\em i.e.} each atom remains in the ground state
along $y$ and $z$. By integrating over these variables, the problem is then
reduced to a one-dimensional two-particle Schr\"odinger equation, with
\be
\label{Ham1D}
{\cal H}^{\beta_1\beta_2}=\sum_{i=1}^2\left[\frac{(p_i)^2}{2m}+
w^{\beta_i}(x_i,t)\right]
+u^{\beta_1\beta_2}_x(x_1-x_2)
\ee
replacing the Hamiltonian (\ref{Hamil}) in Eq.~(\ref{HamilTot}). Here 
$w^{\beta}(x,t)$ is a combination of the $v^{\beta}(x)$ whose form changes with
time, and 
$u^{\beta_1\beta_2}_x$ is an effective interaction potential taking into account
the integration over $y$ and $z$, and therefore depending on the shape of
$v_\perp$. We shall study the dynamics at $t\geq 0$ for different values 
of $(\beta_1,\beta_2)$ separately.
If $\beta_1=\beta_2\equiv\beta$ the total initial 
normalized state, symmetric under particle interchange, is 
\begin{equation}
|\psi^{\beta \beta}(0)\rangle\approx
\frac{|\psi_+\rangle_1|\psi_-\rangle_2+|\psi_-\rangle_1|\psi_+\rangle_2}{\sqrt{2}}
\otimes|\beta\rangle_1\otimes|\beta\rangle_2,
\end{equation}
where the initial overlap 
$\langle\psi_-|\psi_+\rangle\ll 1$ 
has been neglected in computing the normalization.
If both particles are in state $\left| a\right\rangle $,
no interaction takes place
and thus the collisional phase $\phi^{aa}=0$. 
Therefore, we shall now consider in more detail the situation in which  
both particles are in state $\left| b\right\rangle $ and thus move  
within the well $v^b$. In the absence of
interaction, after an oscillation period $T$ they would come
back exactly to the initial state.
Due to the interaction, two effects arise: an additional phase, 
which is accumulated by the wave function as
the number of undergone oscillations increases; and a slight decrease in their 
frequency, because the atoms acquire a small delay in their motion inside 
the trap as they come out from a collision. These effects have to be evaluated in detail,
since they influence the attainable fidelity for a quantum gate based on this scheme.
For symmetry reasons, the relative coordinate motion decouples from the center 
of mass motion, which is not affected by the interaction and can
be solved analytically. For an explicit calculation, it is  now needed to specify
the form of the potentials in Eq.~(\ref{Switch1part}).

\subsubsection{Switching harmonic potentials}

\begin{figure}[tbp]
\begin{center}
\epsfig{file=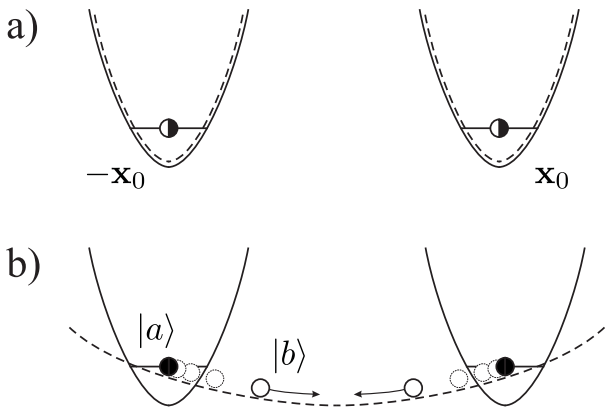,width=8cm}
\vspace*{0.5truecm}
\caption{Configuration at times $t<0$, $t>\tau$ (a) and at $0\leq t\leq\tau$ (b).
The solid (dashed) curves show the potentials for particles in the internal
state $|a\rangle$ ($|b\rangle$). 
\label{FIGswitching}}
\end{center}
\end{figure}

In order to perform the calculations analytically, the potentials in
Eq.~(\ref{Switch1part}) are chosen to be harmonic:
\begin{mathletters}
\bea
v^a(x)&=&\frac{m\omega_0^2}2(x-x_0)^2,\\
v^b(x)&=&\frac{m\omega^2}2x^2,\\
v_\perp(y)&=&\frac{m\omega_\perp^2}2y^2,\label{vperp}
\eea
\end{mathletters}
where $\omega_\perp\gg\omega_0>\omega$. 
Our scheme for gate operation is as follows: initially the two particles are separately
stored in two displaced harmonic wells at $\pm x_0$ as described above, {\em i.e.}
with the potential (Fig.~\ref{FIGswitching}a)
\begin{mathletters}
\bea
w^a(x,t<0)&=&\sum_{\varsigma=+,-}
\Theta(\varsigma x)v^a(\varsigma x),\\
w^b(x,t<0)&=&w^a(x,t<0)
\eea
in the one-dimensional Hamiltonian Eq.~(\ref{Ham1D}).
At $t=0$ the potential undergoes a sudden change, 
namely the barrier between the two wells 
is selectively switched off for state $|b\rangle$ only 
(Fig.~\ref{FIGswitching}b): \end{mathletters}
\begin{mathletters}
\bea
w^a(x,0<t<\tau)&=&w^a(x,t<0);\\
w^b(x,0<t<\tau)&=&v^b(x).
\eea
Then at $t=\tau$, the potential barrier is suddenly restored: 
$w^{a,b}(x,t>\tau)=w^{a,b}(x,t<0)$.
\end{mathletters}
The time evolution at $0<t<\tau$ is characterized by oscillations 
with periodicity $T\equiv 2\pi/\omega$.
The projection of the evolved CM wave function on the initial one 
\begin{equation}
\label{OverPsiCM}
\left| \left\langle \psi^{bb}_{\rm CM}(t)\right| \left. \psi^{bb}_{{\rm CM}%
}(0)\right\rangle \right| ^2=\left[ 1+\frac{\left(\omega _0^2-\omega^2\right)^2}
{4\omega_0^2\omega^2}\sin^2\left( \omega t\right) \right] ^{-\frac 12}
\end{equation}
has instead a period of $T/2$, because of the parity of the spatial
wave function.
The time-dependent energy shift (\ref{deltaE})
due to the interaction turns out to be
\begin{equation}
\Delta E^{bb}(t)=a_s^{bb}\hbar\omega_\perp
\sqrt{\frac{8m\Omega (t)}{\pi \hbar }}
e^{-\frac{2m\omega_0}\hbar x_0^2
\left[1-\sin^2(\omega t)\frac{\omega_0\Omega(t)}{\omega^2}\right]}\;,
\end{equation}
where $\Omega (t)=\omega ^2\omega _0/[\omega ^2\cos^2(\omega t)+
\omega _0^2\sin^2(\omega t)]$.
Hence the interaction-induced phase shift (\ref{phicol}) accumulated after each 
oscillation period $T$ is (evaluating the integral in a saddle-point
approximation)
\begin{eqnarray}
\label{phiPert} 
\phi^{bb}_T &=&\int_0^T\frac{\Delta E^{bb}(t)}\hbar dt
\nonumber\\
&\approx &8a_s^{bb}\sqrt{\frac{m\omega _0}\hbar\frac{\omega _y\omega _z}
{\omega_0^2+\omega^2(4x_0^2\,m\omega_0/\hbar-1)}}\;.  
\end{eqnarray}
If the particles are in different internal states, the center of mass 
does not decouple from the relative motion.
No analytical solution is found in this case, 
and one must resort to numerical techniques to evaluate the collisional phase
$\phi^{ab}$.

\subsubsection{Implementation of conditional dynamics}
\label{SECgatemicrotrap}

If at time $\tau$ the atoms have come back to their initial spatial distribution,
corresponding to a symmetrized product of the ground states of the two wells, 
then after the
barrier is raised they will remain trapped around the original position. The
only change in the overall state will be a phase 
$\phi^{\beta_1}_\tau+\phi^{\beta_2}_\tau+\phi^{\beta_1\beta_2}_\tau$ as
discussed in Sect.~\ref{sect:phinterlattice}.
Therefore,  
the gate operation time $\tau$ has to be chosen in such a way as to maximize the
overlap $|\langle\psi^{\beta_1\beta_2}(\tau)
|\psi^{\beta_1\beta_2}(0)\rangle|^2$ for all $\beta_1,\beta_2$. 
If the modifications in the atomic motion due to interaction are not too strong,
this condition will be satisfied to a good approximation after an integer 
number $n$ of oscillations. Thus, for $\tau\approx nT$, the following mapping
is realized:
\bea
|a\rangle_1 |a\rangle_2 &\rightarrow& 
  e^{-i(2\phi^a_\tau+\phi^{aa}_\tau)} |a\rangle_1 |a\rangle_2,\nonumber\\
|a\rangle_1 |b\rangle_2 &\rightarrow& 
  e^{-i(\phi^a_\tau+\phi^b_\tau+\phi^{ab}_\tau)} |a\rangle_1 |b\rangle_2,\nonumber\\
|b\rangle_1 |a\rangle_2 &\rightarrow& 
  e^{-i(\phi^b_\tau+\phi^a_\tau+\phi^{ba}_\tau)} |b\rangle_1 |a\rangle_2,\nonumber\\
|b\rangle_1 |b\rangle_2 &\rightarrow& 
  e^{-i(2\phi^b_\tau+\phi^{bb}_\tau)} |b\rangle_1 |b\rangle_2,\label{mappingG}
\eea
where $\phi^{aa}_\tau=0$ as discussed in Sect.~\ref{section:phinter}.
If we apply a further single-bit rotation $|0\rangle\langle 0|e^{-i\phi^a_\tau}+
|1\rangle\langle 1|e^{-i(\phi^b_\tau+\phi^{ab}_\tau)}$ (where 
the logical states are defined as $|0\rangle\equiv|a\rangle$ and
$|1\rangle\equiv|b\rangle$) and take into account that  
for symmetry reasons $\phi^{ab}_\tau=\phi^{ba}_\tau$, the mapping
Eq.~(\ref{mappingG}) realizes the fundamental phase gate
\bea
|0\rangle|0\rangle &\rightarrow&|0\rangle|0\rangle,\nonumber\\
|0\rangle|1\rangle &\rightarrow&|0\rangle|1\rangle,\nonumber\\
|1\rangle|0\rangle&\rightarrow&|1\rangle|0\rangle,\nonumber\\
|1\rangle|1\rangle&\rightarrow&e^{-in(\phi^{bb}_\tau-2\phi^{ab}_\tau)}|1\rangle|1\rangle,
\eea
where the phase difference $\phi^{bb}_\tau-2\phi^{ab}_\tau$ has to be adjusted to
$\pm\pi$ by a proper choice of the trap parameters.

\section{Physical realizations}
\label{sect:realizations}

A physical implementation of the scenarios described in Sec.~\ref{EntviaCCC}
requires an interaction which produces internal-state-dependent conservative 
trap potentials and the possibility of manipulating these potentials
independently. The choice of the internal atomic states
$|a\rangle$ and $|b\rangle$ has to be such that they are elastic
(i.e. the internal states do not change after the collision).
To achieve entanglement operations with high fidelity, one has to be able to load 
or cool the atoms to the ground states of the trapping potentials. 
Finally, for the application of parallel quantum computing one needs periodic structures 
(e.g.~optical lattices), together with the ability to control the positions of the atoms
and to fill the lattice sites selectively. 

\subsection{Two-qubit gates in optical lattices}
\label{PhysImpLatt}

In this Section we want to discuss how a number of difficulties can be overcome that
one encounters when trying to use optical lattices for quantum computing. 
We will first show how one can achieve a filling factor of $1$ with 
particles in the ground states (lowest band) of the lattice. This can be achieved 
by using an ultracold 
very dense sample of weakly interacting atoms, namely a Bose-Einstein condensate, 
and slowly turning on an optical potential. The repulsive interaction between
the particles increases as the optical potential is made deeper. At the same time 
the hopping rate at which particles move from one site to the next decreases. 
If the optical lattice is turned on on a time scale much slower than the
hopping rate and the temperature $kT$ can be kept much smaller than the interaction
energy between two particles in one site, one can achieve a filling of the optical 
lattice with exactly one particle per lattice site. \cite{mott} Finally, we note that a filling factor of 
one out of two lattice sites has been achieved in very recent optical lattice experiments.  \cite{opticallattices4}

We will also discuss how the lattice potentials can be moved
in a state-selective way for implementing the two-qubit gate \cite{jaksch98}.
For alkali atoms with a nuclear spin equal to $3/2$ we show how atoms in 
different hyperfine levels can be moved into different directions.
It is clear that other difficulties like e.g.\ addressing single qubits  
exist, but they will not be discussed here since their experimental solution is not 
specific to the present implementation.

\subsubsection{Hamiltonian for a Bose-Einstein condensate in an optical lattice}

We assume a Bose-Einstein condensate of atoms in internal state $|a\rangle$ to be loaded into an optical lattice potential
$V_T({\bf x})+V_0({\bf x})$, where 
\be
V_0({\bf x})=V_{x0}\sin ^{2}(kx)+V_{y0}\sin ^{2}(ky)+V_{z0}\sin ^{2}(kz)
\label{potentialeq}
\ee
is a periodic optical lattice potential and $V_T({\bf x})$ is a superlattice potential 
slowly varying in space compared to $V_0({\bf x})$. $k$ is the wave number of the lasers
producing the lattice potential.
The Hamiltonian reads \cite{mott}
\begin{eqnarray}
H &=&\int d^{3}x\psi ^{\dagger }({\bf x})\left( -\frac{\hbar ^{2}}{2m}\nabla
^{2}+V_{0}({\bf x})+V_{T}({\bf x}) -\mu \right) \psi ({\bf x})  \label{H} \nonumber \\
&&+\frac{1}{2}
\frac{4\pi  a^{aa}_{s} \hbar^2} {m}
\int d^{3}x\psi ^{\dagger }({\bf x})\psi
^{\dagger }({\bf x}) \psi (%
{\bf x})\psi ({\bf x}),
\end{eqnarray}
where $\psi ({\bf x})$ is the bosonic field operator and $\mu$ is the chemical
potential i.e.~a Lagrangian multiplier to fix the number of particles. Expanding the 
field operators in the Wannier basis while keeping only the dominant terms \cite{mott}
Eq.~(\ref{H}) reduces to the Bose-Hubbard Hamiltonian 
\begin{equation}
H=-J\sum_{<i,j>}b_{i}^{\dagger }b_{j}+\sum_{i}\left(\epsilon _{i}-\mu\right)\hat{n}_{i} \, +
\frac{1}{2}U\sum_{i}\hat{n}_{i}(\hat{n}_{i}-1),  \label{BH}
\end{equation}
where the operators $\hat{n}_{i}=b_{i}^{\dagger }b_{i}$ count the number of
bosonic atoms at lattice site $i$; the
annihilation and creation operators $b_{i}$ and $b_{i}^{\dagger }$ obey the
canonical commutation relations $[b_{i},b_{j}^{\dagger }]=\delta _{ij}$.
$J$ is the tunneling matrix element and $U$ describes the (repulsive) 
interaction between particles at the same lattice site. $\epsilon _{i}=V_T(x_i)$ is
the value of the slowly varying superlattice potential at site $i$.
The ratio $U/J$ is controlled by the depth of the optical lattice potential
$V_{j0}$. Increasing $V_{j0}$ (via the intensity of the trapping lasers) reduces the 
tunneling matrix element $J$ and increases the repulsive interaction between the 
atoms $U$ \cite{mott}.

\subsubsection{Loading the lattice}

In order to perform gate operations in optical lattices we have to be able to
selectively fill the lattice sites with exactly one particle. This can be achieved
by making use of the phase transition from a superfluid BEC phase 
to a Mott insulator (MI) phase at low temperatures, which can be induced by 
increasing the ratio of the onsite interaction $U$ to the tunneling matrix 
element $J$ predicted by the Bose-Hubbard model \cite{BH,Bruder}. 
In the MI phase the density $\rho_i$ (occupation number per site) is pinned
at integer $n=0,1,2,\ldots $ corresponding to a commensurate filling of the
lattice, and thus represents an {\it optical crystal }
with diagonal long range order with period imposed by the laser light. 
Particle number fluctuations are thereby drastically reduced
and thus the number of particles per lattice site is fixed. The number of particles per 
lattice site depends on the chemical potential
$\mu$ in the isotropic case $\epsilon_i=0$ \cite{BH}. In the non-isotropic case we may
view $\mu-\epsilon_i$ as a local chemical potential. Therefore $\rho_i$ can 
be controlled by the superlattice potential $V_T({\bf x})$.

\begin{figure}[tbp]
\begin{center}
\epsfig{file=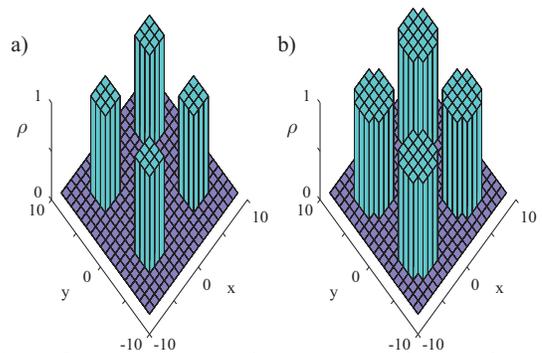,width=7cm}
\caption{Superlattice potential in 2D with $V_{T}(x,y)=40J\;(\sin ^{2}(
\protect\pi x/9a)+\sin ^{2}(\protect\pi y/9a))$ with the spacing between the
lattice sites. The particle density 
$\protect\rho (x,y)$ for four superlattice wells is shown.  
Parameters: a) $U=30J$ and $\protect\mu =15J$, 
b) $U=50J$ and $\protect\mu =27J$.
\label{suplatt}}
\end{center}
\end{figure}

Using a Gutzwiller ansatz \cite{mott,Gutz,GutzI} for the wave function we have performed
a mean field calculation to demonstrate how, by a proper choice of the potential $V_T({\bf x})$, 
one can fill certain blocks of the optical lattice with exactly one particle
at temperature $T=0$. Figure \ref{suplatt} shows the result of this mean field 
calculation, a MI phase where the lattice sites are either filled with $0$ or $1$ particles. 
The number fluctuations are almost equal to zero and thus not shown in this
plot. To achieve a MI phase at finite temperature $T \neq 0$ one has to fulfill the
requirement $kT \ll U$ where the interaction strength $U$ gives the order of magnitude
of the first excitation energy in a MI phase. One also has to ensure that particles do not
move from a filled site with energy $\epsilon_i$ to an adjacent empty site with energy
$\epsilon_j$ i.e.~the temperature has to be much smaller than the
energy difference between these two sites $kT \ll \epsilon_j-\epsilon_i$.
In Sec. \ref{SECqc}, we will need periodic fillings of optical lattices as shown in 
Fig.~\ref{suplatt} to implement efficient multi-particle entanglement operations and 
for parallel quantum computing.

\subsubsection{Moving the lattice potentials state selectively}
\label{MovLatt}

\begin{figure}[tbp]
\begin{center}
\epsfig{file=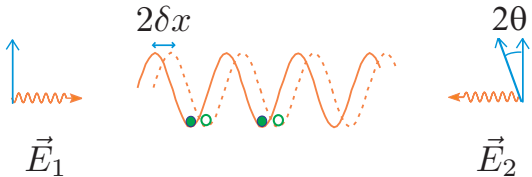,width=7cm}
\caption{Laser configuration along the $z$-axis.
\label{FIGsetup}}
\end{center}
\end{figure}

\begin{figure}[tbp]
\begin{center}
\epsfig{file=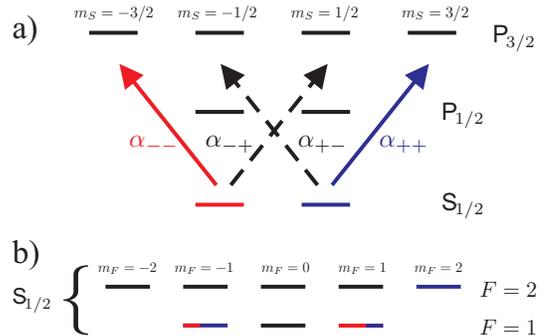,width=7cm}
\caption{Level scheme of $^{87}$Rb and $^{23}$Na and laser configuration. 
(a) Fine structure energy levels and laser configuration. The detuning is
chosen such that the polarizabilities $\alpha_{+-}$ and $\alpha_{-+}$ vanish.
(b) Hyperfine level structure.
\label{Hypstates}}
\end{center}
\end{figure}

We consider the example of alkali atoms with a nuclear 
spin equal to $3/2$ ($^{87}$Rb, $^{23}$Na) trapped by standing waves in three
dimensions and thus confined by a potential of the shape as given in Eq.~(\ref{potentialeq}). The internal states of interest are hyperfine levels
corresponding to the ground state $S_{1/2}$ as shown in Fig.~\ref{Hypstates}b. 
Along the $z$ axis, the standing waves are in the lin$\angle$lin configuration (two linearly
polarized counter-propagating traveling waves with the electric fields $\vec E_1$ and $\vec E_2$
forming an angle $2\theta$ \cite{Finkelstein}) as shown in Fig.~\ref{FIGsetup}. 
The total electric field is a
superposition of right and left circularly polarized standing waves
($\sigma^\pm$) which can be shifted with respect to each other by
changing $\theta$,
\be
\vec E^+(z,t) = E_0 e^{-i\nu t} \left[ \vec \epsilon_+ \sin(kz\!+\!\theta) + 
\vec \epsilon_- \sin(kz\!-\!\theta)\right],
\ee
where $\vec \epsilon_\pm$ denote unit right and left circular polarization vectors,
$k=\nu/c$ is the laser wave vector and $E_0$ the amplitude. The lasers
are tuned between the $P_{1/2}$ and $P_{3/2}$ levels so that the dynamical 
polarizabilities $\alpha_{\pm \mp}$ of the two fine structure $S_{1/2}$ states corresponding to
$m_s=\pm 1/2$ due to the laser polarization $\sigma^{\mp}$ vanish ($\alpha_{+-}=
\alpha_{-+}=0$), whereas
the polarizabilities $\alpha_{\pm \pm}$ due to $\sigma^\pm$ are identical 
($\alpha_{++}=\alpha_{-} \equiv \alpha$). This configuration
is shown in Fig.~\ref{Hypstates}a and can be achieved by tuning the lasers between the
$P_{3/2}$ and $P_{1/2}$ fine state levels so that the ac-Stark shifts of these two levels cancel
each other. The optical potentials for these two states are
$V_{m_s=\pm 1/2}(z,\theta) =  \alpha |E_0|^2 \sin^2\left(kz\pm\theta\right)$.

We choose for the states $|a\rangle$ and $|b\rangle$ the hyperfine structure states
$|a\rangle\equiv|F=1,m_f=1\rangle$ and $|b\rangle\equiv|F=2,m_f=2\rangle$.
Due to angular momentum conservation, these states are stable under collisions (for
the dominant central electronic interaction \cite{Julienne,Tiesinga}). The potentials
``seen'' by the atoms in these internal states are
\begin{mathletters}
\label{Vab}
\bea
V^a(z,\theta) &=& \left[ V_{m_s=1/2}(z,\theta) + 3V_{m_s=-1/2}(z,\theta)\right]/4  \\
V^b(z,\theta) &=& V_{m_s=1/2}(z,\theta).
\eea
\end{mathletters}
If one stores atoms in these potentials and they are deep enough,
there is no tunneling to neighboring wells and we can approximate
them by harmonic potentials. By varying the angle $\theta$ from $\pi/2$ to $0$,
the potentials $V^b$ and $V^a$ move in opposite directions until they completely
overlap. Then, going back to $\theta=\pi/2$ the potentials return to their original
positions. The shape of the potential $V^a$ changes as it moves.
By choosing $\theta(t)=\pi \left(1-\left(1+\exp(-(\tau_i/\tau_r)^2)\right)/
\left(1+\exp((t^2-\tau_i^2)/\tau_r^2)\right)\right)/2$ with $\tau_r=25/\omega$ and 
$\tau_i=25/\omega$, the frequencies and displacements of the harmonic
potentials approximating (\ref{Vab}) are exactly those plotted in Fig.~\ref{Fidel}a.

\subsubsection{Gate fidelity}

\begin{figure}[tbp]
\epsfig{file=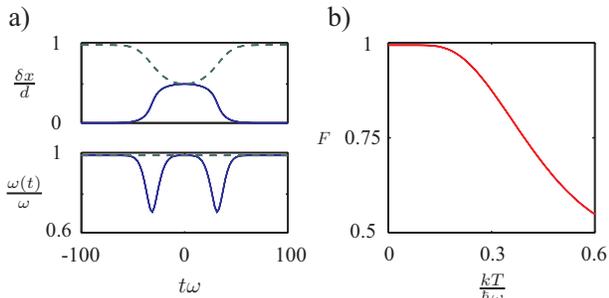,width=8cm}
\caption{a) Upper plot: Displacements $\delta x^a(t)/d$ (solid line) and
$1+\delta x^b(t)/d$ 
(dashed line). Lower plot: Trap frequencies $\omega^{a}(t)/\omega$ 
(solid line) and $\omega^{b}(t)/\omega$ (dashed line). b) Fidelity $F$ against 
temperature $kT/\hbar \omega$ for $^{87}$Rb with $a_s=5.1 {\rm nm}$.
Here $\omega=2 \pi \times 100 {\rm kHz}$ and $d=390 {\rm nm}$.
\label{Fidel}}
\end{figure}

We use the minimum fidelity $F$ \cite{Schu} to characterize the quality of the gate.
$F$ is defined as
\be
\label{FSchu}
F = \min_{\varphi} \langle \tilde\varphi| {\rm tr}_{\rm ext}\left(
{\cal U} |\varphi\rangle\langle\varphi|\otimes \rho_{\rm ext} 
{\cal U}^\dagger \right)|\tilde\varphi\rangle,
\ee
where $|\varphi\rangle$ is an arbitrary internal state of both atoms,
$|\tilde \varphi\rangle$ is the state resulting from $|\varphi\rangle$ using
the mapping (\ref{gate}). The trace is taken over motional states,
${\cal U}$ is the evolution operator for the internal states
coupled to the external motion (including the collision), and $\rho_{\rm
ext}$ is the density operator corresponding to both atoms being at a temperature
$T$ at time $t=-\tau$ \cite{jaksch98}. In Fig.~\ref{Fidel}b the fidelity $F$ is plotted 
as a function of the temperature $T$ for the displacements and trap frequencies shown in
Fig.~\ref{Fidel}a. This figure shows that one can achieve very high fidelities
in realistic situations.

\subsection{Two-qubit gates in magnetic microtraps}
\label{Twogatemagn}

We now consider the implementation of a switching potential by means of
electromagnetic trapping forces. We first discuss the possibility of obtaining 
the desired state dependence by assuming some improvements on devices 
which are now experimentally available \cite{hinds1,hinds2,hinds3}. 
Then we compute the performance of a
quantum gate for realistic trapping parameters.

\subsubsection{Microscopic electromagnetic trapping potential}
\label{sect:mirror}

The interaction between the magnetic dipole moment of an atom in some 
hyperfine state $|F,m_F\rangle$ and an external static magnetic field $\vec B$
entails an energy $U_{\rm magn}\approx g_F\mu_Bm_F|\vec B|$ 
depending on the atomic internal state via the quantum number $m_F$ (here
$\mu_B$ is the Bohr magneton and $g_F$ is the Land\'e factor). 
On the other hand, the Stark shift induced on an atom by an electric field 
$\vec E$ gives a state-independent energy 
$U_{\rm el}\approx\frac 12\alpha |\vec E|^2$, where
$\alpha$ is the atomic polarizability. The interplay between these two effects
can be exploited in order to obtain a trapping potential whose shape depends 
on the atomic internal state.
As an example, we consider an atomic mirror
with an external magnetic field \cite{hinds1,hinds2,hinds3}, providing
confinement along two directions with trapping frequencies which can range from
a few tens of kHz up to some MHz. Microscopic electrodes can be plugged on the
mirror's surface \cite{Schmiedmayer}, thus allowing for the design of a 
potential with the characteristics described in Sect.~\ref{section:switching}.

\subsubsection{Loading and moving atoms within the trap}

Several schemes of loading atoms into the trap have been envisaged
(see for example \cite{hinds1,hinds2,hinds3}). Most of them rely on an intermediate stage 
where atoms can be
trapped and cooled without coming in contact with the magnetic mirror. This 
pre-loading trap can be either initially displaced from the surface, or 
close to it but based on a different trapping mechanism (for instance an 
evanescent wave mirror, where different internal states can be trapped by 
gravity \cite{GOST} before the atoms can be put in the right states for magnetic
trapping), to be replaced by 
the electromagnetic microtrap with a gradual switch-on of the electric and bias
magnetic fields in the final stage of loading \cite{Schmiedmayer}. This could
also allow for implementing a controlled filling of the trap sites, in a similar
way to that already discussed in Sec.~\ref{PhysImpLatt}.
A further feature to be implemented in view of performing more complex 
algorithms is the arrangement of several gate potentials in a periodic pattern,
and the possibility of transporting atoms within this structure. An example
would be given by two adjacent rows of potential minima, shiftable 
with respect to each other, where atoms could be loaded. 
A system like the one suggested in Sect.~\ref{sect:mirror} could allow in
principle to obtain such a configuration, since the magnetic field minima can be
shifted parallel to the surface by rotating the bias magnetic field. In this way
it should be possible to move some atoms, while holding others in place by means
of additional local electric fields \cite{hinds1,hinds2,hinds3}. Provided that atoms can
be addressed individually, which is needed even for performing a one-bit
quantum gate, a procedure for implementing a simple quantum algorithm could be the
following: perform a gate between two suitably chosen atoms, being
close to each other but belonging to different rows,
then mutually displace the rows and select another pair of atoms,
including one of those coming out from the previous gate. Repeat until the
algorithm has been operated, applying the required one-qubit rotations in 
between the above steps and possibly performing some of them in parallel.

\subsubsection{Switching the trap potentials state selectively}

\begin{figure}
\epsfig{figure=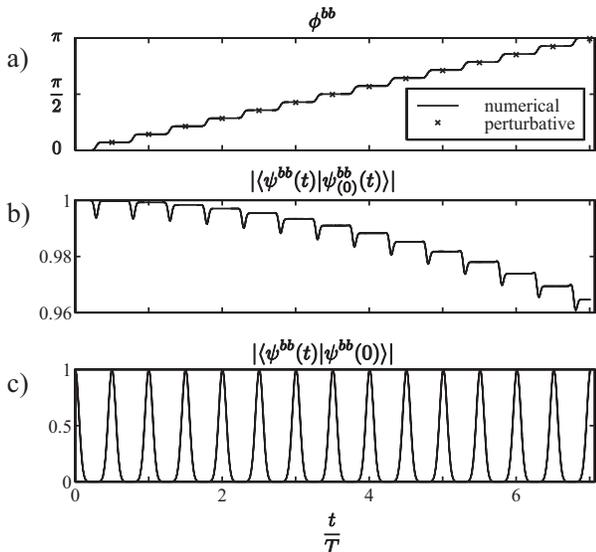,width=8truecm}
\caption{\label{Magnbb}
Dynamics during gate operation, with
both atoms in state $\left| b\right\rangle $: a) interaction-induced phase 
shift - the circles refer to the perturbative calculation 
(\protect\ref{phiPert}); b) projection of the evolved state on the corresponding 
state evolved without interaction; c) projection of the evolved state on the 
initial one.
We choose $\omega\approx 23.4$ kHz and $\omega _y=\omega _z=150$ kHz,
corresponding to ground-state widths $a_x\approx 50$ nm, $a_y=a_z\approx 28$
nm, with the initial wells having frequency $\omega _0=2\omega $ and
displaced by $x_0=3\sqrt{2}a_x$. 
We take for the scattering length the known 
value for $^{87}$Rb, {\em i.e.} $a_s^{bb}\approx a_s^{ab}=5.1$ nm.
Time is in units of the oscillation period $T$.
}
\end{figure}

We choose for the states $|a\rangle$ and $|b\rangle$ the same hyperfine
structure states of $^{87}$Rb considered in the
previous Section, which are low-magnetic field seekers. 
If both particles are in state $|a\rangle$, there is no interaction-induced 
phase shift, as already discussed in Sec.~\ref{section:phinter}.
The results for both particles in state $|b\rangle$ 
are shown in Fig.~\ref{Magnbb}. 
The time dependence of $\phi^{bb}$ is step-like (Fig.~\ref{Magnbb}a): the
collisional phase is incremented at times $t_k\equiv(2k-1)T/4$, 
when the atoms meet at the center of the well, and remains constant at
intermediate times, when they separate again. The influence of the interaction 
on the atomic motion can be seen from Fig.~\ref{Magnbb}b,
depicting the overlap between the evolved interacting two-atom state 
$|\psi^{bb}(t)\rangle$ and the corresponding state $|\psi^{bb}_{(0)}(t)\rangle$ 
computed without taking into account the interaction. The curve has local minima
at times $t_k$, signalling that a collision is taking place, and shows a global
decrease corresponding to a slight delay of the interacting motion with respect
to the non-interacting one.
As it can be seen from Fig.~\ref{Magnbb}c, this effect is not dramatic:
the oscillation period in the presence of
interaction is increased just by $\delta T\approx 2\times 10^{-3}T$ (with the
parameters used here), and
the harmonic potential ensures
that the system comes periodically back to its initial state.
After $7$ oscillations 
we get a phase shift due to the interaction of $\pi $, 
whereas the perturbative formula (\ref{phiPert}) gives $%
7\phi^{bb}_T\approx 0.98\pi$. 
Therefore we choose $\tau=7(T+\delta T)\approx 0.15$ms: 
the overlap between the initial and the evolved wave function at that time is
$|\langle\psi^{bb}(\tau)|\psi^{bb}(0)\rangle|^2\approx 0.996$. 
The behavior turns out to be quite different \cite{CJCZ}
when the atoms are in different internal states:
the phase shift increases more rapidly, but after a few oscillations the
system does no longer come back to the initial state. This has a
simple explanation. The two atoms collide as soon as the one being in state $%
\left| b\right\rangle $, moving within the potential $m\omega ^2x^2/2$,
reaches its turning point, where the other atom is trapped. The interaction
time is therefore longer than if both atoms were in state $\left|
b\right\rangle $. Indeed, in that case they meet at the trap center, with
their maximal velocity. This explains why the system picks up a bigger
phase shift per oscillation period in the present case. On the other hand,
the collision excites the motion of the atom in state $\left| a\right\rangle 
$ within its own well, and therefore the initial state is no longer
recovered. This problem can be avoided if the potential minimum for state $%
\left| a\right\rangle $ is displaced along the transverse direction from the
one for state $\left| b\right\rangle $ by means of an additional electrostatic 
field \cite{hinds1,hinds2,hinds3}, so that the atoms interact if
and only if they are both in state $\left| b\right\rangle $. This problem would not 
exist in an adiabatic scheme for the gate operation, when the shape of the potential
is changed slowly with respect to the atomic motion. This will be the subject of
future investigation.

\subsubsection{Gate fidelity}

The calculation of the fidelity in this case has to take into account the
symmetrization of the wave function under particle interchange, expressed by an
operator $S$ to be explicitly inserted into Eq.~(\ref{FSchu}):
\be
F = \min_{\varphi} \left\{{\rm tr}_{\rm ext}\left[\langle \tilde\varphi| 
{\cal U}S \left(|\varphi\rangle\langle\varphi|\otimes \rho_{\rm ext} \right)
S^\dagger{\cal U}^\dagger |\tilde\varphi\rangle\right]\right\},
\ee
With the parameters quoted
above, we obtain $F > 0.98$. In order to reach such a fidelity, 
timing has to be quite precise, with a
resolution of the order of $10^{-3}T$ corresponding to tens of ns
in this case.

\section{Parallel quantum computing}
\label{SECqc}

In this Section, we will discuss how quantum gates based on controlled
collisions can be exploited for quantum computing. It is clear that, with
the realization of a universal two-bit gate,\ any quantum computation can be
performed, just as it is the case with other implementations. On the other hand, 
manipulations such as moving and switching potentials offer a great deal of 
parallelism \cite{parallelqc1,parallelqc2} not available in other systems.  

We will focus our attention on implementations in optical lattices. 
Some of the ideas could readily be translated into arrays of magnetic
microtraps, if the distances between the individual potential wells could be
made much shorter than present-day state-of-the-art of nanofabrication. In 
such a situation, adiabatic variants of the switching operations  (see comment at
the beginning of Sec.~\ref{section:switching})
can be used to create multi-particle entangled clusters 
of neighboring atoms, similar as with moving potentials. Details of this analysis
will be presented somewhere else \cite{CJCZ}.

\medskip
One may ask, what can be done in optical lattices that cannot be done in other 
implementations? The answer to this question depends on a number of experimental 
conditions such as the possibility of creating  regular filling structures and, like in 
ion-traps, on the possibility of addressing single atoms individually.
In the following, we will first  (Sec.~\ref{SECmulti})  give an example of what can be done with controlled 
lattice movements in conventional set-ups i.e.~with random filling of the lattice sites and without 
any control of the position of individual atoms. We will see that this already allows one to perform  interesting spectroscopic studies of the degree of entanglement between the atoms thus created.
  Next (Sec.~\ref{SECerror}), we will describe  what can be done if one achieves a regular occupation 
of the lattice sites and can address the atoms individually. Under such circumstances, an efficient implementation of quantum error correction and of a quantum memory (concatenated Shor 
code) is possible. Furthermore, fault tolerant versions of certain quantum gates and of  quantum error correction can be
implemented straightforwardly, as will be sketched in (Sec.~\ref{SECfault}). 
  Finally, in Sec.~\ref{SECsweep}, we describe how auxiliary atomic levels can be used to realize highly selective entanglement operations, where individually selected atoms are swept across the lattice to create 
GHZ states \cite{GHZ} of a large number of particles. Together with \ref{SECerror} and \ref{SECfault}, this scheme has all 
the ingredients that are necessary for an efficient realization of fault-tolerant quantum computing. 

\subsection{Multi-particle entanglement operations}
\label{SECmulti}

\begin{figure}[tbp]
\epsfig{file=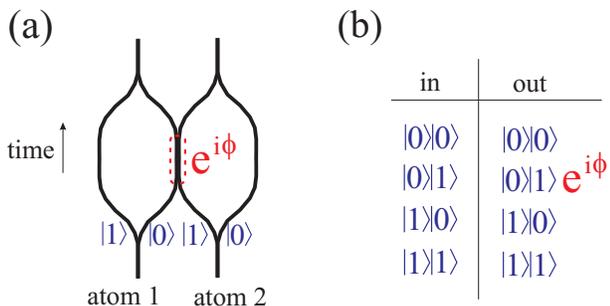,width=8cm}
\caption{Atom-interferometric process realizing the quantum gate. (a) Two-particle interferometer; (b) Truth table.}
\label{FIGgatter}
\end{figure}

The two-qubit gates described in Sec.~\ref{sect:realizations} correspond abstractly
to an atom interferometer as shown in Fig.~\ref{FIGgatter}. The
interferometer has two inputs which are the two atoms trapped at neighboring
potential wells. By shifting the potentials back and forth as described in
Secs.~\ref{Movpot}, only one combination of paths of the two particles overlaps
and leads to a phase shift, namely the paths corresponding to state 
$|a\rangle _{1}$ for the left particle and $|b\rangle _{2}$ for the right
particle. To emphasize the role of the internal states as {\em logical}
states, we shall henceforth use the notation $|0\rangle \equiv |a\rangle $
and $|1\rangle \equiv |b\rangle $ and neglect the kinetic phases $\phi^a$, $\phi^b$ as they
appear in (\ref{gate}). Furthermore, we drop the atomic index as long as
there is no danger of confusion.  

The logical truth table 
corresponding to the interferometric process is shown in Fig.~\ref{FIGgatter}.
[A similar identification of logical states can be made in magnetic traps as is pointed out in
Sec.~\ref{SECgatemicrotrap}. The labelling of the paths for the left particle in the interferometer has to be 
interchanged in this case.] For $\phi =\phi ^{01}=\pi $ this realizes a phase gate \cite{cirac95}. 
The phase gate and the set of all one-bit unitary transformations, which can be realized by Raman laser pulses on 
the internal states $|0\rangle $ and $|1\rangle $, define a {\em universal set of quantum gates.} 
\cite{universalgate1,universalgate2,universalgate3,universalgate4}

\begin{figure}[tbp]
\epsfig{file=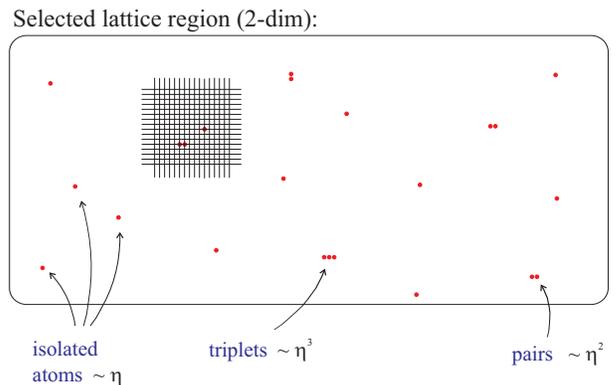,width=8cm}
\caption{Random occupation of a two dimensional lattice with single atoms.}
\label{FIGrandom}
\end{figure}

\begin{figure}[tbp]
\epsfig{file=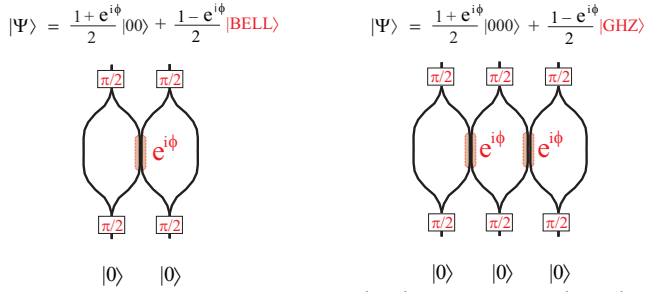,width=8.5cm}
\caption{Entanglement of pairs (left) and triplets (right) of neighboring
atoms by a single lattice shift.}
\label{FIGbellghz}
\end{figure}

An important difference between optical lattices and other implementations
is given by the global effect of the lattice manipulations. To illustrate
this point, consider first a two dimensional lattice as in Fig.~\ref{FIGrandom}
with random occupation of the sites and a filling factor $\eta \ll 1$, where 
$\eta $ is defined as the average number of atoms per lattice site.
Let us assume that the loading of the lattice can be accomplished in such a way that there are 
no multiply occupied lattice sites, i.e.~that each lattice site is occupied by no more 
than a single atom. 
\noindent Then, in any region of the lattice, one will find isolated atoms, pairs of
neighboring atoms, triplets, and so forth, with a relative frequency
proportional to $\eta $, $\eta ^{2}$, $\eta ^{3},$ respectively. Consider
now the following {\em Ramsey experiment} \cite{Ramsey} where initially all atoms are 
prepared in the internal state $|0\rangle $ and in the motional ground state of their
individual potential wells. In some selected region of the lattice, the
following sequence of operations is applied: (1) a $\pi /2$ laser pulse
brings all atoms into a superposition of the internal states $|0\rangle 
$ and $|1\rangle $; (2) the lattice is shifted across one lattice site and
then, after a variable length of time, shifted back to its original
position, (3) finally a second $\pi /2$ pulse is applied to the region. The
effect of this sequence is illustrated in Fig.~\ref{FIGbellghz}. For a
group of $N=1,2,3,\dots $ neighboring atoms, the lattice shift corresponds
to a {\em N-particle interferometric process.} 
\noindent Specifically, one obtains the following transformations. For isolated atoms: 
\begin{equation}
|0\rangle \longrightarrow |0\rangle ;
\end{equation}
for pairs of neighboring atoms: 
\begin{equation}
|00\rangle \longrightarrow \frac{1+e^{i\phi }}{2}|00\rangle +\frac{%
1-e^{i\phi }}{2}|\mbox{\sc bell}\rangle \,;  \label{paare}
\end{equation}
and for triplets of neighboring atoms: 
\begin{equation}
|000\rangle \longrightarrow \frac{1+e^{i\phi }}{2}|000\rangle +\frac{%
1-e^{i\phi }}{2}|\mbox{\sc ghz}\rangle \,;  \label{tripletts}
\end{equation}
where we have used the notation 
\begin{eqnarray}
|\mbox{\sc bell}\rangle &=&\frac{1}{\sqrt{2}}\left\{ |0\rangle |+\rangle
-|1\rangle |-\rangle \right\} \,,  \nonumber \\
|\mbox{\sc ghz}\rangle &=&\frac{1}{\sqrt{2}}\left\{ |0\rangle |+\rangle
|1\rangle -|1\rangle |-\rangle |0\rangle \right\} \,,
\end{eqnarray}
and $|\pm \rangle =(|0\rangle \pm |1\rangle )/\sqrt{2}$. The expressions for
groups of more particles become more complicated and shall be ignored in
the present discussion. It is clear that for $\phi =\pi $ Bell- and GHZ
states \cite{Bell,GHZ} are created by a {\em single} lattice shift at various places within
the region. This corresponds to an ensemble of 2-bit and 3-bit 
quantum gates, respectively, acting simultaneously at different lattice sites. 
To analyze the states (\ref{paare}) 
and (\ref{tripletts}) spectroscopically one could measure the state of the
atoms in a final step of the above Ramsey sequence e.g.\ by a fluorescence
measurement. It is clear that by such a measurement the entangled states
will be destroyed. On the other hand, by repeating this sequence many times
with different samples, one can measure the fluorescence signal as function
of the phase $\phi $ (interaction time). Under ideal circumstances, all
isolated atoms will remain in the dark state $|0\rangle $ while all
fluorescence signals come from Bell ($\sim \eta ^{2}$) or GHZ ($\sim \eta
^{3}$) states \cite{footnote2}. To check that entangled states,
rather than mixtures, are created in the process, the experiment is
performed with different interaction times, e.g.\ times corresponding to 
$\phi =\pi $ and $\phi =2\pi $. For entangled states as in (\ref{paare}) and (
\ref{tripletts}) all fluorescence signals will vanish at $\phi =2\pi $, while
this will not be the case if the states created by the atomic collisions are
mixtures of classical many-particle states. More generally, by measuring the 
{\em visibility} of the fluorescence signal one may study the {\em fidelity
of the entanglement} created in the process, and its dependence on certain
noise sources such as a finite temperature of the atoms. This way, the curve 
plotted in Fig.~\ref{Fidel}b) could be tested experimentally.

\subsection{Quantum error correction}
\label{SECerror}

\begin{figure}[tbp]
\begin{center}
\epsfig{file=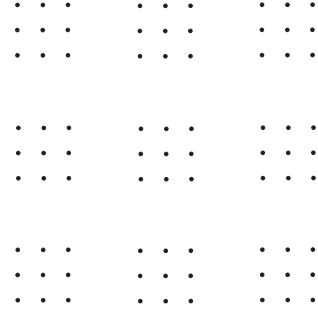,width=4cm}
\vspace{1em}
\caption{Ordered arrangement of atoms in an optical lattice (see also Fig.~\ref{suplatt}).}
\label{FIGparallel}
\end{center}
\end{figure}

\begin{figure}[tph]
\epsfig{file=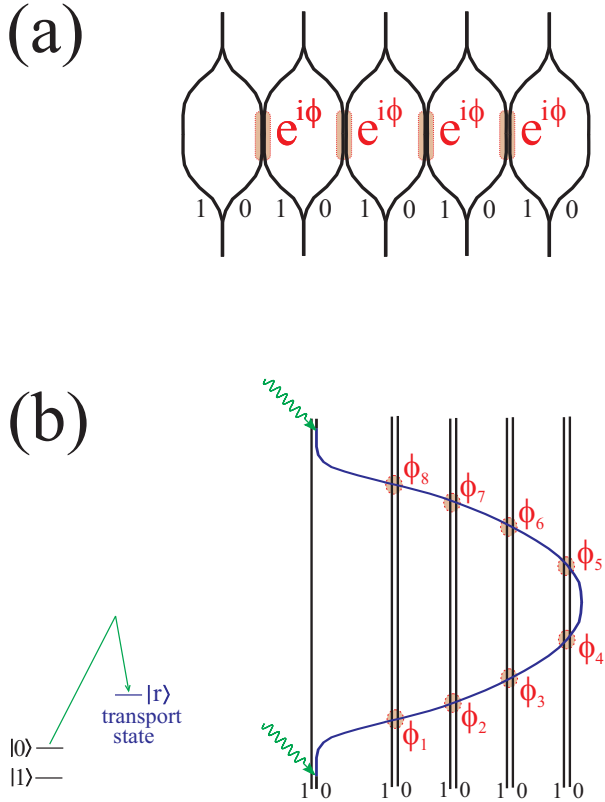,width=8cm}
\caption{(a) ``Shift operation'': The internal atomic states $|0\rangle $ and 
$|1\rangle $ couple to different lattice potentials that are moved against each other
as explained in Sec. \ref{MovLatt}. This corresponds to a multi-particle interferometer where the 
same phase shift is acquired whenever two paths temporary overlap.
By simple lattice manipulations, therefore, entire groups of atoms become entangled.
(b) ``Sweep operation": For more selective entanglement operations, a third atomic level 
 $|r\rangle $ is used {\protect  \cite{foot3}}. In this scheme, only atoms in the level 
$|r\rangle $ are moved, whereas the states $|0\rangle $ and $|1\rangle $ are kept in the same 
potential. At the beginning of an entanglement operation, the atoms are first excited from one 
of the states $|0\rangle $ or $|1\rangle $ to the state $|r\rangle$ before the lattice is
moved. This scheme is much more selective in the sense that those atoms
which shall participate in a gate operation are first activated, before they
couple to the moving lattice, and the collisional phases $\phi_j$ can be varied for each 
interaction individually.}
\label{FIGshift_sw}
\end{figure}

To employ these entanglement operations for quantum computing, one has to have
precise control over the number and the location of atoms that are involved
in the collisional process. In addition to the ability of addressing single
atoms, one therefore has to achieve a certain {\em ordered occupation of the
lattice sites}. As described in Sec. III.A., Fig.~\ref{suplatt}, this can be by achieved by
controlling the intensity of the trapping laser at sufficiently low temperatures. This way 
optical crystals with periodic patterns of atoms can be created as indicated in Figs. \ref
{FIGparallel} and \ref{FIGselfsim} \cite{weitz-priv}
Under such circumstances the parallelism of the lattice manipulations can be
exploited advantageously. On one side, similar logical operations can be performed
simultaneously at different locations on the lattice. On the other side, as
we have seen in Fig.~\ref{FIGbellghz}, a single lattice shift can entangle
whole groups of atoms. Two types of such entanglement operations are shown
in Fig.~\ref{FIGshift_sw}. One involves only the logical states $|0\rangle $
and $|1\rangle ,$ while the second uses a third atomic level as a
``transport state'' (see Sec.~\ref{SECsweep}), into which any atom must first be 
activated, before it can participate in an entanglement operation. In the following,
we will first discuss applications of the shift operation as in Fig.~\ref{FIGshift_sw}(a).
Later, in Sec.~\ref{SECsweep} we will consider a more flexible (``sweep'') operation
shown in Fig.~\ref{FIGshift_sw}(b).

An application of shift operations as described in Fig.~\ref{FIGshift_sw}(a) concerns the 
realization of a {\em  quantum memory}, where a qubit $\alpha |0\rangle +\beta |1\rangle $ (with
unknown coefficients $\alpha $ and $\beta $) is encoded in the quantum state
of a larger block of atoms and stabilized against decoherence with the help
of quantum error correction \cite{shor-code,QECC1,QECC2,laflamme96,QECC3,QECC4,QECC5,QECC-review}. A 
particular quantum code that is able to protect a qubit
against general 1-bit errors (spin flip and phase flip) has been
proposed by Shor \cite{shor-code}. It is a 9-bit code where the codewords 
\begin{eqnarray}
|0_{{\rm S}}\rangle &=&2^{-3/2}(|000\rangle +|111\rangle )(|000\rangle
+|111\rangle )(|000\rangle +|111\rangle )  \nonumber \\
|1_{{\rm S}}\rangle &=&2^{-3/2}(|000\rangle -|111\rangle )(|000\rangle
-|111\rangle )(|000\rangle -|111\rangle )\nonumber \\
 &&  \label{EQshorcode}
\end{eqnarray}
consist of products of certain GHZ states. Abstractly speaking, the encoding
operation consists of a mapping (embedding) of the qubit's 2-dimensional
Hilbert space ${\cal H}$ into a $2^{9}$-dimensional Hilbert space of the form 
\begin{equation}
{\cal H}\ni \alpha |0\rangle +\beta |1\rangle \mapsto \alpha |0_{{\rm S}%
}\rangle +\beta |1_{{\rm S}}\rangle \in {\cal H}_{{\rm E}}\subset {\cal H}%
^{\otimes 9}.  \label{shorencode}
\end{equation}
To stabilize the encoded information against decoherence, the code must be
measured and corrected on a time scale $\tau \ll 1/9\gamma $ where $\gamma $
is the rate of decoherence for a single qubit. This is possible since all
errors that may occur on any one of the qubits of the codewords (\ref
{EQshorcode}) map the code into a family of 2-dimensional subspaces of $%
{\cal H}^{\otimes 9}$ which are all orthogonal on ${\cal H}_{{\rm E}}$
\cite{shor-code}.

\begin{figure}[tph]
\epsfig{file=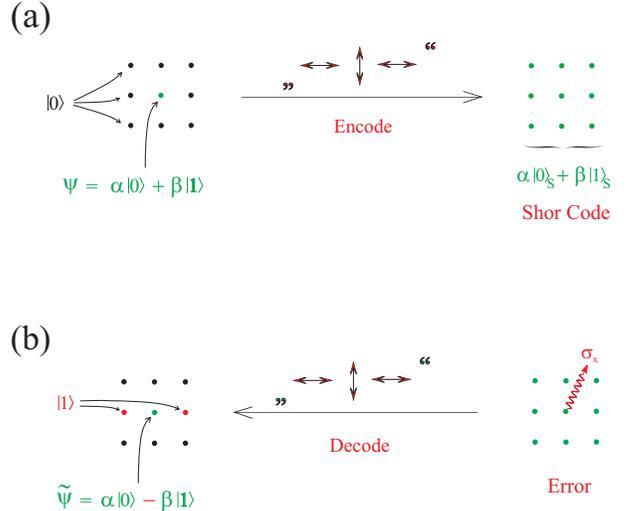,width=8cm}
\caption{(a) Encoding of a qubit into a block of $3\times 3$ atoms; (b)
Decoding and syndrome measurement.}
\label{FIGshor12}
\end{figure}

The Shor code (\ref{EQshorcode}) can be implemented efficiently in a two
dimensional lattice configuration \cite{freedman98} as in Fig.~\ref{FIGparallel}, by using the
shift operation of Fig.~\ref{FIGshift_sw}(a). \ To see this, imagine that
the qubit/atom whose state is to be encoded is surrounded by neighboring
atoms as in Fig.~\ref{FIGshor12}. The idea is of course to encode the
central qubit in the whole block of $3\times 3$ \ qubits. Initially the
central atom is in the unknown state $|\psi \rangle =\alpha |0\rangle +\beta
|1\rangle $ while all neighboring atoms are in state $|0\rangle $. As is
shown in Appendix B, the initial state is transformed into the Shor code
by a simple sequence of horizontal and vertical lattice shifts combined with
certain 1-bit rotations, as indicated in Fig.~\ref{FIGshor12}(a). \ By this
process, the information contained in $\psi $ is so to speak de-localized
over the whole block of 9 atoms. To check whether an error has occurred on
one of the qubits, the block is first decoded
by \ the inverse transformation  \cite{laflamme96}, which involves the same sequence of lattice
shifts as the encoding. Subsequently, one measures which of the neighboring
atoms are in the state $|1\rangle $. 
In the language of quantum error correction, the surrounding atoms of the
central qubit in Fig.~\ref{FIGshor12}(b) are the carriers of the {\em error
syndrome} \cite{laflamme96}, meaning that their state gives information regarding what type of
error occurred and, more importantly, which unitary 1-bit rotation has to be
applied to the central qubit to restore it to the original state. In a
fluorescence measurement, this information corresponds to a specific pattern
of bright and dark atoms surrounding the central qubit. For example, in 
Fig.~\ref{FIGshor12}(b) a spin flip has occurred in the central qubit. This means that the state of
the {\em block} is transformed into $\alpha \sigma _{x}^{{\rm ctr}}|0_{{\rm S%
}}\rangle +\beta \sigma _{x}^{{\rm ctr}}|1_{{\rm S}}\rangle $ where $\sigma
_{x}^{{\rm ctr}}$ is the corresponding Pauli spin operator associated with
the central atom. By the decoding operation of Fig.~\ref{FIGshor12}(b), this
state is mapped into a product state of 9 qubits in which the neighboring
atoms in the central row are both in the (bright) state $|1\rangle ,$ while
the other surrounding atoms are in the (dark) state $|0\rangle .$ The
central atom is in a state equivalent to $|\psi \rangle $ up to a unitary 
transformation. Similar fluorescence patterns are obtained if a phase error 
occurs in the central atom or an arbitrary 1-bit error on any of the atoms in
the block. A complete table of the error syndrome is given in the Appendix.

\begin{figure}[tph]
\epsfig{file=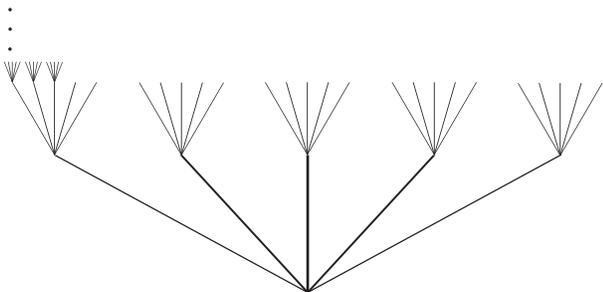,width=8cm}
\caption{Concatenated quantum coding. At each coding level, a single qubit is encoded in 
a block of a larger number of qubits, here 5. (See e.g. \protect\cite{fault-tolerant-review}).}
\label{FIGconcat}
\end{figure}

\begin{figure}[tph]
\begin{center}
\epsfig{file=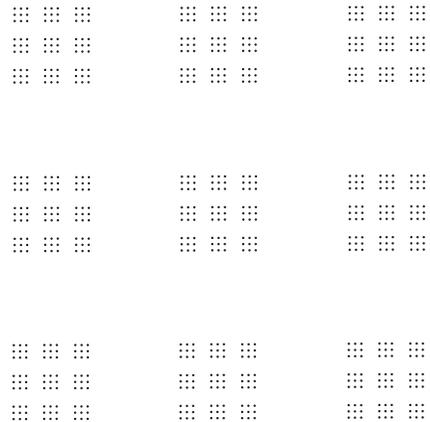,width=5.5cm}
\caption{Concatenated quantum coding in an optical lattice. At each coding level, 
two-dimensional lattice displacements with an increasing periodicity are applied.
The nested character of Fig.~\ref{FIGconcat} is here reflected by a self-similar filling 
pattern of the lattice.}
\label{FIGselfsim}
\end{center}
\end{figure}

The essential point is that the measurement on the
surrounding atoms does not reveal nor destroy the state of the central atoms
(the coefficients $\alpha $ and $\beta $ remain unknown throughout the
process). If the sequence of operations ``decode-correct-encode'' is
repeated sufficiently often within the decoherence time $1/9\gamma $ of the
block, the state $\psi $ may be protected over arbitrarily long times, in
principle. 
Here one assumes, of course, that the decoding and encoding operations 
themselves are {\em free
of errors}. In our situation this means that all phases acquired in the
atomic collisions can be perfectly controlled. Since these operations will
always bear some imperfection/imprecision, the probability that an error is
introduced by an imperfect operation increases/accumulates with repeated
applications of these operations. The general solution to this
quantum-memory problem was given by Knill and Laflamme \cite{knill96} and by
others \cite{othersthanknill1,othersthanknill2,othersthanknill3}, and 
requires a {\em concatenation of
encoding operations} as shown schematically in Fig.~\ref{FIGconcat}. 
The number of required concatenation steps depends on how long the qubit is
to be stored. It can be shown \cite{knill96} that, given the precision of
the operations is above a certain threshold, a qubit can be stored for an
arbitrary long time, where the number of qubits required for encoding (i.e.
the length of the code) grows polynomially with the length of the storage
time. In the optical lattice configuration, a concatenation of the encoding
can be implemented straightforwardly. Imagine that, in the central block in
Fig.~\ref{FIGparallel}, the center atom is initially in state $|\psi \rangle
=\alpha |0\rangle +\beta |1\rangle $ (similar as in Fig.~\ref{FIGshor12})
whereas all other atoms are in state $|0\rangle $. This means that both the
surrounding atoms in the center block and the atoms of all the other
blocks are initially in state $|0\rangle $. The first step of the
encoding operation is identical as in Fig.~\ref{FIGshor12} and results in
the configuration where the center block is in a superposition of the Shor
code words $|0_{{\rm S}}\rangle $ and $|1_{{\rm S}}\rangle ,$ whereas the
surrounding blocks remain in state $|0\rangle $. In the second step, the
same operation is repeated on a larger scale, i.e. the lattice is shifted
across a larger distance such as to make the {\em blocks} temporarily
overlap while the 1-bit operations of the first step are now repeated on
corresponding atoms of the outer blocks. As a result, the information $|\psi
\rangle $ originally carried by the center atom is now delocalized over 
$9\times 9=81$ atoms! This scheme may be iterated as indicated in Fig.~\ref
{FIGselfsim}. 
When in the second (and higher-order) encoding step the blocks are brought
to overlap, one has to make sure that only phases between corresponding atoms
of the different blocks are accumulated. The most elegant way to achieve
this would be with the aid of a technique where the 0 and 1 states are
displaced vertically before the atoms are moved. This could be implemented 
in a three-dimensional lattice configuration\cite{footnote3}. The shift operation
is then really a ``lift \& shift'' operation. The collisional interaction
is then only switched on by varying the vertical displacement, after
the blocks have been moved horizontally. If such a lifting technique can not
be implemented, e.g. in a truly two-dimensional configuration, then during the horizontal 
motion there will be also collisions between non-corresponding atoms, for example the 
atoms in the right column of one of the blocks with atoms in the left column of a
neighboring block. To avoid these unwanted phase shifts, it is possible to
vary the velocity of the lattice movement in such a way that during unwanted
collisions a phase of $e^{2\pi i}$ is acquired. This method is clearly more
susceptible to decoherence. On the other hand, our numerical studies have shown 
\cite{jaksch98}, that by an appropriate choice of the displacement function $%
\theta (t)$ in Fig.~\ref{FIGsetup}, the phase of a single collision
can, in principle, be controlled with a very high precision (with fidelity $\geq 0.9997$)
and the probability for exciting phonons remains correspondingly small \cite{footnote4}.

It does not seem impossible that $\theta(t)$ could be controlled 
precisely enough to meet the threshold of fault-tolerant computation \cite{fault-tolerant-review}, 
but we have not yet made detailed numerical investigations for this situation. 
In summary, the method of {\em concatenated coding} can be implemented in optical 
lattices by repeated sequences of lattice displacements on {\em self-similar filling structures}.

\bigskip

\subsection{Fault-tolerant computing}
\label{SECfault}

In a quantum computer, we do not only wish to store quantum information, but
also to process it in a quantum algorithm. To prevent an accumulation of errors
during the calculation due to imperfect gate operations, one needs to use
fault-tolerant quantum gates that act on the encoded information.
Furthermore, errors should be corrected fault tolerantly, that is, without
decoding the information (and therefore exposing the qubit to decoherence).
The general theory of fault-tolerant computation has been developed by
several researchers \cite{fault-tolerant-review}. In optical lattices,
many of such fault-tolerant operations have a geometrically intuitive
implementation. For example, if two qubits are encoded in blocks of 9 atoms
each, as in Fig.~\ref{FIGshor12}, a controlled-NOT operation can be
implemented by moving one block on top of the other so that each pair of
corresponding atoms from the two blocks share a single potential well and
acquire a phase shift $e^{i\pi }.$ [This is a straightforward generalization of the
situation in Fig.~\ref{FIGgatter}]. 
\begin{figure}[tph]
\epsfig{file=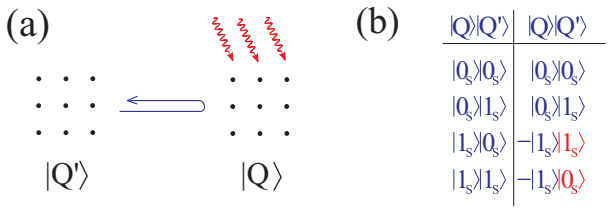,width=8cm}
\caption{Implementation of a fault-tolerant CNOT gate.}
\label{FIGcnot}
\end{figure}
\noindent When a $\pi /2$ pulse is applied on one of the blocks before and after the
blocks are shifted, a fault-tolerant realization of the CNOT gate, with a
truth table as in Fig.~\ref{FIGcnot} is realized. The minus sign may be
eliminated by applying a $3\pi /2$ pulse instead of the second $\pi /2$
pulse. Similarly, one can find a simple fault-tolerant realization of the NOT gate,
while for example the Hadamard transform is more involved and requires a
measurement with auxiliary qubits. Whether or not one can find similarly
{\em efficient} implementations for a complete set of fault tolerant gates, is
still under investigation.

\begin{figure}[tbp]
\begin{center}
\epsfig{file=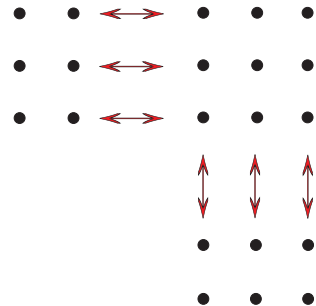,width=4cm}
\caption{Implementation of fault-tolerant error correction.}
\label{FIGarmada}
\end{center}
\end{figure}

To check whether an error has occurred during a gate operation, one has to
measure whether the blocks are still in a superposition of the correct
codewords. For the Shor code, this can be done in the following way \cite
{shor-code}, see Fig.~\ref{FIGarmada}: To detect a spin-flip, one has to
measure the parity of the first two atoms in any row and compare it to the
parity of the last two atoms of every row. To do this one would use an
``Armada'' of $3\times 2$ ancillas in the state $(|00\rangle +|11\rangle
)(|00\rangle +|11\rangle )(|00\rangle +|11\rangle )$, which approaches the
block from the left in Fig.~\ref{FIGarmada} by moving the lattice horizontally. 
\noindent To measure the parities, the Armada is moved on top of the first two columns
of the data block so that the atoms interact pairwise with atoms of the data block and acquire 
a phase shift of $e^{i\pi}$. To satisfy the criteria for fault tolerance, we need to avoid
collisions while the ancillas are moved on top of the code, and thus need a
``lift \& shift'' implementation of the operation, as mentioned earlier.  
Suppose there was a spin-flip in one of the atoms of the first row.
Then the state of the ancillary atoms after the interaction reads $(-|00\rangle
+|11\rangle )(|00\rangle +|11\rangle )(|00\rangle +|11\rangle )$, and the
error will be detected by measuring the parity of  the ancillas in
each row, after applying a Hadamard transform. In a second run, the Armada is reset
in the initial state and then is moved on top of  the last two rows of the block, and so
on. To detect a phase-flip, a similar procedure is used with an Armada of $%
2\times 3$ atoms that approaches the block in Fig.~\ref{FIGarmada} from {\em below} by 
moving the lattice vertically. Since these ancillas
should measure any change of sign in any of the GHZ states that make up the
codewords (\ref{EQshorcode}), \ they have to be prepared in the state $%
|000000\rangle +|111111\rangle.$  A phase flip can then be detected as
previously, where now a Hadamard transform has to be applied to the block
first, before the ``attack'' starts from below. In the specific implementation using
optical lattices, one could also think about other schemes using only a single
row of ancilla atoms on each side of the data block in Fig.~\ref{FIGarmada} as realized
in Fig.~\ref{suplatt}b).

\subsection{Selectivity and ``sweep operations''\label{SECselectivity}}
\label{SECsweep}

The examples discussed so far make use of the parallelism of the lattice
shift to implement certain multi-particle entanglement (or gate) operations
efficiently. On the other hand, the shift operation as described in Fig.~\ref
{FIGshift_sw}(a) is too rigid, when certain operations should apply to a
selected group of atoms only. This problem can in principle be solved by
using a third atomic level $|r\rangle $ as indicated in Fig.~\ref
{FIGshift_sw}(b). In this scheme, the level $|r\rangle $ couples dominantly
to a transport lattice \cite{foot3}, while the ``logical states'' $|0\rangle $ and $%
|1\rangle $ are kept in the same potential. At the beginning of an
entanglement operation, the atoms are first excited from one of the states $%
|0\rangle $ or $|1\rangle $ to the state $|r\rangle ,$ before the lattice is
moved. This scheme is much more selective in the sense that those atoms
which shall participate in a gate operation are first activated, before they
can participate in the lattice movement. All operations that we have
discussed can then be realized in the same manner, with the additional
property that only those atoms, to which the operation $|1\rangle
\rightarrow |{\rm r}\rangle $ is applied, will participate. With this additional
feature, it is clear, that {\em universal computations} can be implemented.

\begin{figure}[tbp]
\epsfig{file=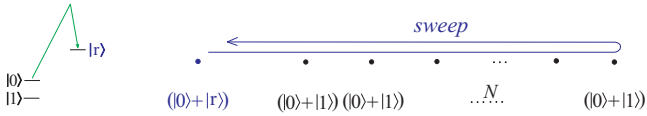,width=8.5cm}
\caption{Realization of an $(N+1)$-dimensional GHZ state by a single sweep
operation.}
\label{FIGn-ghz}
\end{figure}

Another merit of this scheme is that one can realize more flexible entanglement operations. 
Consider, for example, a 1-dimensional situation as in Fig.~\ref{FIGn-ghz} 
with a string of $N$\ atoms initially prepared in the product state $%
(|0\rangle +|1\rangle )^{\otimes N}$ and a selected additional atom (left)
in the state $(|0\rangle +|{\rm r\rangle )}$. By moving the transport
lattice, the selected atom is swept across the $N$ lattice sites. During
that motion, it interacts with each of the $N$ atoms thereby transforming the
state of each atom into $e^{{i\phi ^{0}}}|0\rangle +e^{i\phi ^{1}}|1\rangle
, $ with a differential phase $\phi =\phi ^{1}-\phi ^{0}.$ The resulting
total state is of the form 
\begin{eqnarray}
|0\rangle (|0\rangle +|1\rangle )(|0\rangle +|1\rangle )\cdots (|0\rangle
+|1\rangle ) &&  \nonumber \\
+|{\rm r}\rangle e^{iN\phi ^{0}}(|0\rangle +e^{i\phi }|1\rangle )(|0\rangle
+e^{i\phi }|1\rangle )\cdots (|0\rangle +e^{i\phi }|1\rangle ). &&
\end{eqnarray}
As long as the collisional phases are different ($\phi \neq 0$) for the two
logical states, $\phi $ can by varied with the speed of this sweep
operation. For $\phi =\pi $ one obtains a $N+1$ dimensional GHZ state (see
Fig.~\ref{FIGn-ghz} ) which can easily be brought to the standard form 
\begin{equation}
|\psi \rangle =\frac{1}{\sqrt{2}}\left( |0\rangle |0\rangle |0\rangle \cdots
|0\rangle +|1\rangle |1\rangle |1\rangle \cdots |1\rangle \right) .
\end{equation}
Note that for the creation of this state only a single sweep
operation is required!

\begin{figure}[tph]
\epsfig{file=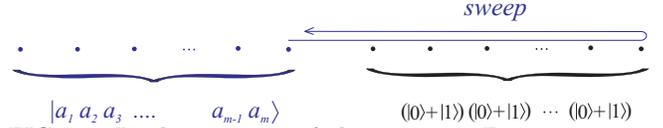,width=8.5cm}
\caption{Implementation of the quantum Fourier transform by a sweep
operation with variable speed.}
\label{FIGqft}
\end{figure}

This scheme can be generalized in several directions. By varying the speed
by which the lattice is moved during the sweep operation, the phases can be
controlled individually for each atom of the string as indicated in Fig.~\ref{FIGshift_sw}, 
allowing for more complex entanglement operations. As a final
example consider a configuration as in Fig.~\ref{FIGqft}(b), with a ``source
register'' consisting of a string of $m$ atoms in the state $|a\rangle
=|a_{1}\;a_{2}\;a_{3}\;\cdots \;a_{m}\rangle $, $a_{j}\in \{0,1\}$ and a
``target register'' of $m$ further atoms in the state $(|0\rangle +|1\rangle
)^{\otimes m}$, similar as in Fig.~\ref{FIGn-ghz}. 
The state vector $|a\rangle =|a_{1}\;a_{2}\;a_{3}\;\cdots \;a_{m}\rangle $
should be interpreted as a binary representation of the number $%
a=a_{1}2^{m-1}+a_{2}2^{m-2}+\dots +a_{m}2^{0}.$ Consider now the following
operation where the source register is first activated to couple to the transport
lattice, meaning that each of the atoms $1$ to $m$ that is in state $%
|1\rangle $ \ is excited to state $|r\rangle $. Next, the lattice is moved
to the right so that atoms of the source and the target register interact;
this motion continues with variable speed until the source register
completely overlaps with the target register. It is helpful to mentally
decompose this operation into discrete steps. In the first step, the
transport lattice is shifted one lattice site to the right such that the $m$th 
atom of the source register interacts with the first atom of the target
register. One can tune the interaction time such that a certain phase shift is
acquired during this interaction, namely $\phi =2\pi /2^{m}$. In the next
step, the transport lattice is moved one lattice site further to the right
such that now the $m$ th atom of the source register interacts with the
{\em second} atom of the target register, while at the same time the $m-1$ th atom
of the source register interacts with the first atom of the target register.
In this step, the interaction time is made double as long as in the first
step, so that $\phi =2\pi /2^{m-1}$, and so on. After the lattice has been
moved across $m$ sites in this vein, the total state of the source and the
target register is given by 
\begin{eqnarray}
|a_{1}\;a_{2}\;a_{3}\; &\cdots &\;a_{m}\rangle \otimes (|0\rangle +e^{2\pi
i0.a_{1}a_{2}\dots a_{m}}|1\rangle )  \nonumber \\
&&(|0\rangle +e^{2\pi i0.a_{2}\dots a_{m}}|1\rangle )  \nonumber \\
&&\quad \cdots \quad (|0\rangle +e^{2\pi i0.a_{m}}|1\rangle )\,.
\end{eqnarray}
Finally, the lattice is shifted back to the original position without
changing the phases any more (modulo $2\pi $, see earlier remark, or the
process can be made symmetric such that only half the phase values are
accumulated during the motion to the right while the second halves of the
phase values are accumulated when the lattice is brought back to its
original position.) The overall effect of this sweep operation can be
summarized in the form 
\begin{equation}
|a\rangle |\quad \rangle \longrightarrow e^{i\Phi (a)}|a\rangle |{\cal F}%
(a)\rangle
\end{equation}
wherein $|a\rangle $ and $|\quad \rangle $ denote the initial state of the
source and the target register, and $|{\cal F}(a)\rangle $ is the discrete 
{\em Fourier transform} of the function $j\rightarrow a_{j}$. The additional
phase factor $e^{i\Phi (a)}$ accounts for a possible phase shift arising from collisions
among different atoms of the source, if no vertical displacement of the transport
lattice is possible. It should be remarked that for a superposition of
different input states the source and the target register become entangled,
and therefore a direct application of this method in the Shor algorithm \cite{shor94,ekert96}
is not possible. Nevertheless, this final example
demonstrates a remarkable  flexibility of the entanglement operations that are
possible in optical lattices and similar systems, offering 
new perspectives for efficient implementations of quantum algorithms.

\section{Final remarks}

It is clear that, at the present time, most of the experimental requirements have yet to
be realized, before one can implement quantum computing. There are, however,
recent achievements in cooling and trapping of atoms in optical lattices and in magnetic 
microtraps which make it seem possible that some of these elements could be 
implemented in the laboratory in the near future. There are short-term and long-term perspectives. 
Essential for all quantum information experiments is a
successful {\em cooling} of the atoms to the ground state of a three
dimensional lattice. Numerical calculations \cite{jaksch98} using realistic parameters give 
$kT<0.2\hbar \omega $ as a critical value. Under these 
circumstances, one could perform interesting Ramsey-type spectroscopic studies of the fidelity
of multi-particle entanglement as discussed earlier. To do this, neither
single-atom addressability is required nor are regular filling structures.
When the latter requirements can be realized, on the other hand, coding experiments 
can be done and a quantum memory be implemented. Finally, if one can find three-level 
schemes with different scattering phases for the logical states, universal
computations can be performed. The parallelism of the lattice could then be
exploited for efficient implementations of fault-tolerant quantum computing.

We have discussed multi-particle entanglement \linebreak schemes mainly in the context
of optical lattice implementations. Some of these ideas could readily be
adopted in implementations with magnetic microtraps if one uses adiabatic
schemes. A basic requirement for this is the possibility of creating quantum
dots that are spatially sufficiently close to each other. These ideas will be
discussed somewhere else \cite{CJCZ}.

\acknowledgments 

We thank E.~Hinds, J.~Schmiedmayer, M. Weitz and T. W. H\"ansch for many useful discussions.
We also thank David DiVincenzo and Andrew Steane for helpful discussions on fault-tolerant 
quantum computing during the Benasque Workshop 1998.
One of us (T.~C.) thanks M.~Traini and S.~Stringari for the kind hospitality at
the Physics Department of Trento University, and the ECT* 
for partial support during the completion of this work. This work was supported in part 
by the \"Osterreichischer Fonds zur F\"orderung der wissenschaftlichen Forschung, 
the European Community under the TMR network ERB-FMRX-CT96-0087, the Institute for 
Quantum Information GmbH, and by the Schwerpunktsprogramm ``Quanten-Informationsverarbeitung'' 
der Deutschen Forschungsgemeinschaft.

\appendix

\section{One particle in a moving harmonic potential}
\label{MovHarm}

\subsection{Hamiltonian}

The center of the potential with frequencies $\omega_x$, $\omega_y$ and $\omega_z$ is 
assumed to be given by $\bar{\bf x}(t)=(\bar x(t),0,0)$ and
the Hamiltonian reads
\begin{equation}
H=H_{x}+H_{y}+H_{z},
\label{moveH}
\end{equation}
where $H_{z}=\hbar \omega_{z}(a_{z}^{\dagger}a_{z}+1/2)$, $H_{y}=\hbar
\omega_{y}(a_{y}^{\dagger}a_{y}+1/2)$, and
\begin{equation}
\label{moveHx}
H_{x}= \hbar \omega_{x}\left(a_{x}^{\dagger}a_{x}+\frac{1}{2}+(a^{\dagger}_{x}+a_{x})
\frac{\bar x(t)}{\sqrt{2}}+\frac{\bar x(t)^2}{2}\right).
\end{equation}
The $a$'s are bosonic destruction operators and $\bar x(t)$ is given in harmonic
oscillator units. We will concentrate on the $x$-direction leave out the subscript
$x$ and normalize energies to $\hbar \omega$.

\subsection{Exact solution}

The Schr{\"o}dinger equation for the Hamiltonian Eq.~\ref{moveHx} can be solved exactly \cite{QNoise,GalPas}.
To do so we define 
\begin{equation}
H_0=a^{\dagger}a+\frac{1}{2},
\end{equation}
\begin{equation}
K(t,-\tau)=\frac{1}{\sqrt{2}} \int_{-\tau}^t ds \, \bar x(s) e^{i(s+\tau)}
\end{equation}
and
\begin{equation}
\beta(t,-\tau)=i\int_{-\tau}^t ds \, \left( K(s,-\tau) \{\partial_s K^*(s,-\tau)\}+
\frac{i \bar x(s)^2}{2} \right).
\end{equation}
If initially at time $-\tau$ the system is in the state 
$|\Psi(-\tau)\rangle=|0\rangle$, where $|n\rangle$ is the $n$-th harmonic
oscillator eigenstate we get
\begin{equation}
|\Psi(t)\rangle=e^{i\beta(t,-\tau)} \sum_n \frac{(iK(t,-\tau)e^{-i(t+\tau)})^n}{\sqrt{n!}}
|n\rangle.
\end{equation}
The kinetic phase $\phi$ is thus given by the phase of the overlap of $|\Psi(t)\rangle$ with
the instantaneous ground state $D({\bf \bar x}(t))|0\rangle$, where
$D(\gamma)=\exp(\gamma a^{\dagger}-\gamma^* a)$ denotes the displacement operator
\begin{equation}
\phi=-{\rm arg}\left(\langle 0|D({\bf \bar x}(t))^\dagger |\Psi(t)\rangle\right).
\label{phiex}
\end{equation}
The interaction phase can be found by Eq.~(\ref{phicol}) with the known $|\Psi(t)\rangle$.

\subsection{Corrections to the adiabatic approximation}

We assume $\bar x(t)$ to be an analytic function of $t$ and 
that $\bar x(t) \gg \partial_t \bar x(t) \gg \partial_t^2
\bar x(t) \gg \dots \gg \partial_t^n
\bar x(t)$. By expanding in orders of the time derivatives 
we can write for $K(t',-\tau)$
\begin{eqnarray}
K(t',-\tau)&=&\frac{1}{\sqrt{2}} \left(
i^{N+1} \int_{-\tau}^{t'} ds \{\partial^{N+1}_s \bar x(s)\} e^{i(s+\tau)} - \right. \nonumber \\
&&\left.\sum_{n=0}^N i^{n+1} \{\partial^n_s \bar x(s)\} e^{i(s+\tau)}|^{s=t'}_{s=-\tau}\right),
\label{adiaHO}
\end{eqnarray}
where $N$ is a positive integer.
Note that if we may neglect all terms of order greater than $\partial_t \bar x(t)$ and start in a coherent state
$|\Psi(-\tau)\rangle=D(\bar x(-\tau)+i\partial_t \bar x(t)|_{t=-\tau})|0\rangle$ the state will always be a coherent
state with $\langle x(t) \rangle=\bar x(t)$ and $\langle p(t) \rangle=\partial_t \bar x(t)$.

Now we assume for simplicity that $\bar x(\tau)=\bar x(-\tau)=0$, 
$(\partial_t \bar x(t))|_{t=-\tau}=(\partial_t \bar x(t))|_{t=\tau}=0$
and
$(\partial^n_t \bar x(t))|_{t=-\tau}=(-1)^n(\partial^n_t \bar x(t))|_{t=\tau}$ for $n>1$. 
The system is assumed to be in the state 
$|\Psi(-\tau)\rangle=|0\rangle$, initially. We keep all the terms to fourth order in
the derivatives (in the integrand) and find
\begin{eqnarray}
K(t,-\tau)&=&\frac{1}{\sqrt{2}}\left(\left\{i (\partial^2_t - \partial^4_t)
\bar x(t)\right\}(e^{i(t+\tau)}-1)- \right.\nonumber \\
&&\left. \{\partial^3_t \bar x(t)\}(e^{i(t+\tau)}+1)
\right),
\label{Kapp}
\end{eqnarray}
and
\begin{eqnarray}
\beta(t,-\tau)&=&
\frac{1}{2} \int_{-\tau}^t ds \left(\{\partial_s \bar x(s)\}^2+
\{\partial^2_s \bar x(s)\}^2\right) \nonumber \\ 
&&+\frac{1}{2}\{\partial^2_s \bar x(t)\}^2
(1-e^{-i(t+\tau)}).
\end{eqnarray}
If we choose $(t+\tau)=2 n \pi$ with integer $n$ the largest correction to the 
approximation to the kinetic phase discussed in Sec.~\ref{KinPhas} is of third order.
Also the amplitude of the first excited state is of third order as can be seen from
Eq.~(\ref{Kapp}).

\section{Quantum error correction and the implementation of Shor's code}

Consider a one-dimensional configuration with a string of $n$ atoms, where  $x_0, x_1, x_2, 
\dots, x_n \in \{0,1\}$ label the internal state of the atoms at position $0,1,2,\dots, n$ of the 
lattice. An elementary lattice-shift operation as given in Fig.~\ref{FIGshift_sw}(a) 
is then described as 
\begin{eqnarray}
LX: &|x_0,x_1,x_2,\dots,x_n\rangle \longmapsto&  \nonumber \\
 & e^{-i\sum_j (x_j+1\,{\rm mod}2)x_{j+1}\varphi_{j+1}} 
|x_0,x_1,x_2,\dots,x_n\rangle &
\label{elem_gate}
\end{eqnarray}
where the phase $\varphi_{j+1}$ in the exponent depends on the interaction time and the 
interaction strength between two atoms at the lattice site $j+1$, and the
addition is performed modulo 2. Note that two neighboring atoms at the sites $j$ and $j+1$ 
contribute to the exponent if and only if $x_j=0$ and $x_{j+1}=1$. The variables $x_j$ can 
only take on the values $0$ and $1$.  In all examples we discuss here, $\varphi_j=\varphi = \rm constant$ and is the same for all lattice sites.

The operation (\ref{elem_gate}) defines a generalized phase gate that acts on a group of $n$ neighboring atoms via shifting the lattice across one lattice site. It can easily be seen that, for example when  combined with $\pi/2$-pulses as in Fig.~\ref{FIGbellghz}, LX produces the entangled states (\ref{paare}) and (\ref{tripletts})  for $n=2$ and $n=3$.
 
In two dimensional lattices, the logical variables $x_{kl}$ are labeled by two indices, where $k$ 
is the horizontal index and $l$ the vertical index.
The phase gates corresponding to horizontal and vertical lattice shifts are then defined as
\begin{eqnarray}
LX|\{x_{kl}\}\rangle &=& e^{-i\sum_j (x_{jl}+1\,{\rm
mod}2)x_{j+1,l}\varphi_{j+1,l}}|\{x_{kl}\}\rangle\nonumber\\
LY|\{x_{kl}\}\rangle &=& e^{-i\sum_j (x_{kj}+1\,{\rm
mod}2)x_{k,j+1}\varphi_{k,j+1}}|\{x_{kl}\}\rangle
\label{LXLY}
\end{eqnarray}
as an obvious generalization of  (\ref{elem_gate}).

It is clear that the operations can be further generalized to lattice shifts across an arbitrary number of lattice sites and along arbitrary directions. There are interesting topological questions in this general situation. For the present discussion, however, the gates LX, LY as defined in (\ref{LXLY}) are sufficient and we will set $\varphi_{k,l}=\pi$. Apart from these gates, we will only need 
single-particle operations, in particular the Pauli-operators 
$\sigma_{x,j}, \sigma_{y,j},\sigma_{z,j}$ and the Hadamard transformation ($\pi/2$ pulse)
$H_j=(\sigma_{z,j}+\sigma_{x,j})/\sqrt{2}$ applied to an atom with index $j$.

Consider now a configuration of  $3\times 3$ atoms as in Fig.~\ref{FIGshor12}, where the central atom is in the unknown state $|\psi\rangle = \alpha|0\rangle+\beta |1\rangle$ while  all surrounding atoms are initially in the state $|0\rangle$.  Let us first look at the special case when $|\psi\rangle = |0\rangle$, that is, the central atom is in the state $|0\rangle$ as well. If we apply a $\pi/2$ pulse to each atom of the block and then the operation LX, we obtain a tensor product of three GHZ states where each row of the block is in the same state $(0+1)0(0-1)-(0-1)1(0+1)$.  
(For notational brevity, we suppress the bracket notation in the following and identify $0\equiv|0\rangle$ and $1\equiv|1\rangle$). This state can be transformed to the form $000-111$ by applying $H_1$ to the first atom and $H_3\sigma_{z,3}$ to the third atom of each row.
The operation LX, supplemented by one-qubit rotations,  produces thus one of the code words in (\ref{EQshorcode}).

To realize a quantum memory, an  {\em unknown} state 
$\psi =\alpha 0 + \beta 1$ of the central atom is to be encoded into an entangled 9-bit 
state as in (\ref{shorencode}). 
Let us write the initial (unencoded) state of the block in the form 
\begin{equation}
 |{\tt bare}\rangle = 0_10_20_3\; 0_4(\alpha 0_5+\beta 1_5)0_6\; 0_70_80_9 
\label{unencoded}
\end{equation}
where the first, second, and third triplet refers to the upper, center, and 
lower row of the block in Fig.~\ref{FIGshor12}. 
To encode  $\psi$ into a corresponding superposition of both codewords, lattice 
movements in both horizontal and in vertical direction are required.
In detail, the {\em encoding operation} is given by 
\begin{equation}
 ENC = H_{46}\circ LX\circ H_{456}\circ LY\circ \sigma_{x;369} 
       H_{134679}\circ LX\circ H^{\rm s}
\label{shor_encode}
\end{equation}
whose essential part is a sequence of three lattice movements, {\tt horizontal - vertical -
horizontal}, with certain 1-bit unitary transformations in between.   
In the notation used here, $H^{\rm s}$ denotes a Hadamard transform applied 
to each of the 8 syndrome atoms, whereas $H_{ijk\dots}$ and 
$\sigma_{z;ijk\dots}$ are single-qubit rotations
applied to the selected atoms $i$, $j$, $k$, $\dots$, only.
Applied to the  state (\ref{unencoded}), $ENC$ produces
\begin{eqnarray}
 ENC\; |{\tt bare}\rangle &=& \alpha
(000-111)(001-110)(000+111)\nonumber\\
                        && \beta (000+111)(100+011)(000-111) \nonumber\\
                        &\equiv & \alpha 0_{\rm L} + \beta  1_{\rm L}\,.
\end{eqnarray} 
The codewords $0_{\rm L}$ and $1_{\rm L}$ are equivalent  to the Shor
code (\ref{EQshorcode}), as we shall see presently.

The {\em decoding operation} is given by the inverse of (\ref{shor_encode}),
\begin{equation}
 DEC = H^{\rm s}\circ LX\circ H_{134679}\sigma_{x;369}  \circ LY\circ 
       H_{456}\circ LX\circ H_{46}
\end{equation}
involving the same lattice movements, but the 1-bit operations carried out 
in reverse order. To see explicitly how one can correct an error occurring on one of the qubits $j=1,2,\dots 9$ , we apply the error operators $\sigma_{x,j}$, $\sigma_{y,j}$, or $\sigma_{z,j}$ 
to the encoded state $\alpha 0_{\rm L} + \beta  1_{\rm L}$. Then we apply the decoding operation 
$DEC$ and measure the state of the syndrome atoms.
The code $0_{\rm L}$ and $1_{\rm L}$ is error correcting if every possible error is mapped into a syndrome state different from $0_10_20_3\; 0_40_6\; 0_70_80_9$, and for each syndrome we can tell which unitary transformation has to be applied to the central qubit to restore it to its 
original state [it is not necessary that all errors are mapped to mutually
orthogonal subspaces \cite{shor-code}]. The following table gives for each error the corresponding syndrome and the state of the central qubit:

\bigskip
\begin{tabular}{|c|c|c|}
error & syndrome & central qubit \\ \hline 
  \rm none & 000\,00\,000 & $\alpha 0 + \beta$ 1\\
  && \\ 
$\sigma_{x,1}$ & 110\,00\,000 & $\alpha 0 - \beta$ 1 \\
$\sigma_{x,2}$ & 101\,00\,000 & $\alpha 0 - \beta$ 1 \\
$\sigma_{x,3}$ & 011\,00\,000 & $\alpha 0 - \beta$ 1 \\
  && \\ 
$\sigma_{x,4}$ & 010\,10\,010 & $\alpha 0 + \beta$ 1 \\
$\sigma_{x,5}$ & 000\,11\,000 & $\alpha 0 - \beta$ 1 \\
$\sigma_{x,6}$ & 010\,01\,010 & $\alpha 0 + \beta$ 1 \\
  && \\ 
$\sigma_{x,7}$ & 000\,00\,110 & $\alpha 0 - \beta$ 1 \\
$\sigma_{x,8}$ & 000\,00\,101 & $\alpha 0 - \beta$ 1 \\
$\sigma_{x,9}$ & 000\,00\,011 & $\alpha 0 - \beta$ 1 \\
  && \\ 
$\sigma_{y,1}$ & 100\,00\,000 & $\alpha 0 - \beta$ 1 \\
$\sigma_{y,2}$ & 111\,00\,000 & $\alpha 0 - \beta$ 1 \\
$\sigma_{y,3}$ & 001\,00\,000 & $\alpha 0 - \beta$ 1 \\
  && \\ 
$\sigma_{y,4}$ & 000\,01\,000 & $\alpha 1 - \beta$ 0 \\
$\sigma_{y,5}$ & 010\,00\,010 & $\alpha 1 - \beta$ 0 \\
$\sigma_{y,6}$ & 000\,10\,000 & $\alpha 1 - \beta$ 0 \\
  && \\ 
$\sigma_{y,7}$ & 000\,00\,100 & $\alpha 0 - \beta$ 1 \\
$\sigma_{y,8}$ & 000\,00\,111 & $\alpha 0 - \beta$ 1 \\
$\sigma_{y,9}$ & 000\,00\,001 & $\alpha 0 - \beta$ 1 \\
  && \\ 
$\sigma_{z,1}$ & 010\,00\,000 & $\alpha 0 + \beta$ 1 \\
$\sigma_{z,2}$ & 010\,00\,000 & $\alpha 0 + \beta$ 1 \\
$\sigma_{z,3}$ & 010\,00\,000 & $\alpha 0 + \beta$ 1 \\
  && \\ 
$\sigma_{z,4}$ & 010\,11\,010 & $\alpha 1 - \beta$ 0 \\
$\sigma_{z,5}$ & 010\,11\,010 & $\alpha 1 + \beta$ 0 \\
$\sigma_{z,6}$ & 010\,11\,010 & $\alpha 1 - \beta$ 0 \\
  && \\ 
$\sigma_{z,7}$ & 000\,00\,010 & $\alpha 0 + \beta$ 1 \\
$\sigma_{z,8}$ & 000\,00\,010 & $\alpha 0 + \beta$ 1 \\
$\sigma_{z,9}$ & 000\,00\,010 & $\alpha 0 + \beta$ 1 \\
\end{tabular}

\bigskip
In Fig.~\ref{FIGshor12}, the syndrome atoms visually encircle the unknown qubit that is to be
protected. If any of the 9 qubit suffers a spin flip, a phase flip, or both,
the error can be detected by measuring the state of the syndrome atoms after
the decoding operation has been applied to the group. This could be done by a fluorescence measurement where atoms in state 1 and 0 correspond to  ``bright'' and ``dark'', respectively. 
For example, according to above table, the pattern 
\begin{center} 
\begin{tabular}{rcl} 
0&0&0  \\
1& $\psi'$ &1  \\
0&0&0  
\end{tabular}
\end{center}
tells us that a spin flip has occurred in the central atom,
whereas 
\begin{center} 
\begin{tabular}{rcl} 
0&0&0  \\
0& $\psi'$ &0  \\
1&1&0  
\end{tabular}
\end{center}
reveals a spin flip in the left atom of the lower row, and 
\begin{center} 
\begin{tabular}{rcl} 
0&1&0  \\
1& $\psi'$ &1  \\
0&1&0  
\end{tabular}
\end{center}
corresponds to a phase flip in any of the atoms of the central row. 
In any case, the state $\psi'$ of the central qubit after the detection 
of an error is related to the initial state $\psi$ via a (known) unitary 
operation $U$: $\psi'=U\psi$, which can be obtained from the third 
column of the syndrome table given above.

The fact that the encoding operation involves only 3 lattice movements
provides a specific example of a ``parallelization  of a quantum circuit" 
\cite{parallelqc1,parallelqc2}. We have not proven that 3 is really 
the minimum number of entanglement operations needed; there might be still faster 
sequences. The original Shor code can be recovered from this code 
by applying an additional vertical lattice shift, LY, and certain 1-bit rotations.

% ===============================================================


\begin{thebibliography}{99}
\bibitem{footnote2}
For small filling factors, the signals coming from different groups of atoms
could even be spatially resolved.
\bibitem{foot3}
In $^{87}$Rb and $^{23}$Na one could use $|a\rangle\!\equiv \!|F=1,m_F\!=\!-1\rangle$,
$|b\rangle\!\equiv \!|F=2,m_F\!=\!2\rangle$ and $|c\rangle\!\equiv
\!|F=\!1\!,m_F\!=\!1\rangle$.
\bibitem{freedman98} 
Freedman and Meyer \cite{parallelqc3} have 
independently identified Shor's code as a two-dimensional topological 
quantum code (see also Kitaevs work \cite{parallelqc4,othersthanknill2}). 
The connection between their results and our lattice implementation remains to be clarified.
\bibitem{footnote3} 
Here we assume that one is able to selectively load a two-dimensional ``sheet'' 
in a three-dimensional  lattice.
\bibitem{footnote4} 
It remains the practical question, how well the polarization $\theta(t)$ 
can be controlled in the experiment.
\end{thebibliography}
\end{document}